\documentclass[a4paper,11pt]{article}
\usepackage{jheppub}

\usepackage{slashed}
\usepackage{lineno}
\usepackage{graphicx}
\usepackage{lipsum}
\usepackage{slashed}
\usepackage{amsmath}
\usepackage[normalem]{ulem}
\usepackage{enumitem}\usepackage{mathtools}
\usepackage{bm}
\usepackage{hyperref}
\usepackage[all]{hypcap}
\usepackage{bbm}
\usepackage{xcolor}
\usepackage{subcaption}
\usepackage{pifont}
\usepackage{multirow}
\usepackage{tablefootnote}
\usepackage{cleveref}
\usepackage{multirow}
\usepackage{latexsym}

\usepackage{array}
\DeclareUnicodeCharacter{2212}{-}
\newcolumntype{L}[1]{>{\raggedright\let\newline\\\arraybackslash\hspace{0pt}}m{#1}}
\newcolumntype{C}[1]{>{\centering\let\newline\\\arraybackslash\hspace{0pt}}m{#1}}
\newcolumntype{R}[1]{>{\raggedleft\let\newline\\\arraybackslash\hspace{0pt}}m{#1}}
\newcommand{\cC}{\mathcal{C}}
\newcommand{\cL}{\mathcal{L}}

\usepackage{tikz}
\usetikzlibrary{calc,tikzmark,fit,shapes.geometric,matrix,decorations.markings,arrows.meta,decorations.pathmorphing,patterns,positioning,snakes,patterns.meta}
\tikzset
  {midarrow/.style={decoration={markings,mark=at position 0.5 with
     {\arrow[thin,xshift=2pt]{Triangle[length=4pt,#1]}}},postaction={decorate}}
  }

\tikzset{
proton/.style = {circle, draw=black, thin, fill=black!20!white, minimum size=#1,
              inner sep=0pt, outer sep=0pt},
proton/.default = 6pt 
}

\tikzset{
blob/.style = {circle, draw=black, thin, preaction={fill, black!20!white}, pattern=north east lines, minimum size=#1,
              inner sep=0pt, outer sep=0pt},
blob/.default = 6pt 
}

\tikzset{
wc/.style = {circle, fill, minimum size=#1,
              inner sep=0pt, outer sep=0pt},
wc/.default = 6pt 
}

\tikzset{vector/.style={decorate, draw=black, decoration={snake=coil,pre length=0, post length = 0}}}

\begin{document}
\preprint{DESY-25-152}

\title{
Could electron-top interactions spoil the measurement of the Higgs trilinear?\\[.1cm]
 {\fontsize{14pt}{0}\selectfont---A quantitative estimate at future lepton colliders---}
}

\author[a]{Lukas Allwicher,} 
\emailAdd{lukas.allwicher@desy.de}
\affiliation[a]{Deutsches Elektronen-Synchrotron DESY, Notkestr. 85, 22607 Hamburg, Germany}

\author[a,b]{Christophe Grojean,} 
\emailAdd{christophe.grojean@desy.de}
\affiliation[b]{Institut f{\"u}r Physik, Humboldt-Universit{\"a}t zu Berlin, 12489 Berlin, Germany}

\author[b]{and Lucine Tabatt}
\emailAdd{lucine.marie.anais.tabatt@hu-berlin.de}

\abstract{
The measurement of the Higgs self-coupling is considered the next milestone in the study of the Higgs boson properties. 
At future $e^+e^-$ facilities below the double Higgs production threshold, this is extracted from the $Zh$ production cross-section, which is sensitive to the trilinear coupling at the one-loop level.
At the same perturbative order, potential effects beyond the Standard Model (SM) may affect the Higgstrahlung rate and distort the self-coupling determination.
We study the question focusing especially on contact interactions containing two electron and two top-quark fields.
We conclude that, in the context of FCC-ee and its planned runs at different energies, $eett$ interactions change the Higgs self-coupling sensitivity below the percent level. Even in the most pessimistic scenarios, we confirm a robust sensitivity of the order of 17\% at the 1$\sigma$ confidence level under the assumption of otherwise SM-like Higgs couplings. A crucial role in these results is played by the measurement of fermion pair production above the $Z$ resonance.
}

\maketitle
\flushbottom

\allowdisplaybreaks

\section{Introduction}

The Standard Model (SM) has been established as a remarkably successful description of nature at the microscopic level, with experimental tests confirming its robustness up to the TeV scale and beyond.
The study of the Higgs boson, as its last-discovered component, is one of the main goals of the current and future experimental programme.
Indeed, many of the couplings of the Higgs to other SM fields are still only loosely bound, with e.g. light family Yukawas being constrained within $\mathcal{O}(100)$ times their SM value.
Among the least-known couplings is also the self-coupling.
These couplings are parts of the SM that are so central to its structure, or to the world around us, that redundant or complementary measurements are welcome independently of their unique sensitivity to New Physics.
Writing the generic potential for the neutral, CP-even component of the Higgs doublet as
\begin{align}
    V(h) = m_h^2 h^2 + \kappa_3 \lambda_3^{\rm SM} v h^3 + \kappa_4\lambda_4^{\rm SM} h^4 +\kappa_n \frac{h^n}{v^n}\,,
\end{align}
with \begin{align}
    \lambda_3^{\rm SM} = \frac{m_h^2}{2v^2}\,, \
    \lambda_4^{\rm SM} = \frac{1}{4} \lambda_3^{\rm SM} \,,
\end{align}
where the vacuum expectation value $v \simeq 246$\,GeV is directly extracted from muon decays ($G_F$),
the SM predicts the Higgs self-couplings in terms of other already measured observables:
\begin{align}
    \kappa_3^{\rm SM} = \kappa_4^{\rm SM} = 1\,, \ \kappa_{n>4}^{SM}=0\,,
\end{align}
and $\kappa_i$ parametrise deviations from the SM in the usual way.
Current bounds on $\kappa_3$ are $-1 \lesssim \kappa_3 \lesssim 7$ and come from di-Higgs production at the LHC~\cite{CMS:2025ngq,ATLAS:2024ish}.
At hadron colliders, an individual extraction of the trilinear is obtained from the measurement of the gluon fusion double Higgs production~\cite{Glover:1987nx, DiMicco:2019ngk}, however differential distributions are rather insensitive to the value of $\kappa_3$ while the total rate might be affected by other BSM effects, like a contact interaction of two top-quarks with two Higgs bosons, that will weaken the $\kappa_3$ determination. At lepton colliders, $\kappa_3$ can  be inferred similarly from the double Higgs production in association with a Z boson or via $WW$ fusion, provided that the collider can reach at least 500\,GeV~\cite{DiMicco:2019ngk}. 
Another interesting possibility  is offered by an high-intensity $e^+e^-$ collider such as the proposed FCC-ee, which gains sensitivity at Next-to-Leading Order (NLO) to the Higgs self-coupling~\cite{McCullough:2013rea,DiVita:2017vrr,Maltoni:2018ttu}.
By running above the $Zh$ threshold (the planned FCC-ee baseline runs are at 240 and 365\,GeV centre-of-mass) an absolute measurement of the $e^+e^-\to Zh$ cross-section promises a precision on the extraction of $\kappa_3$ at the $\mathcal{O}(20\%)$ level \footnote {This refers to the combination of HL-LHC and FCC-ee in a global fit~\cite{FCC:2025lpp,deBlas:2025gyz}.}.
However, several other BSM effects can also potentially enter the same Higgstrahlung process, possibly spoiling the $\kappa_3$ precision too.
In particular, it has been pointed out~\cite{Bellafronte:2025ubi} that four-fermion contact interaction involving electrons and tops may, due to colour- and Yukawa-enhanced effects, lead to a modification of the Higgsstrahlung process.
Such interactions would obviously come with a very rich phenomenology. The question we want to tackle in this work is therefore which precision we can expect from other related processes in order to tame this contamination.

The remainder of the paper is structured as follows.
In the next two sections we review NLO effects in the $Zh$ cross section at lepton colliders within the SMEFT and interplays between the Higgs trilinear and other possible BSM effects.
Section~\ref{sec:eettpheno} details the phenomenology of electron-top interactions in processes other than the Higgstrahlung, while Section~\ref{sec:SMEFT} discusses the implications for the extraction of the Higgs self-coupling.
Before concluding, we discuss similar phenomenology in the context of simplified models featuring leptoquark extensions of the SM in Section~\ref{sec:models}.

\section{The Higgs Trilinear in SMEFT}

\begin{figure}
    \centering
    \includegraphics[width=0.75
    \linewidth]{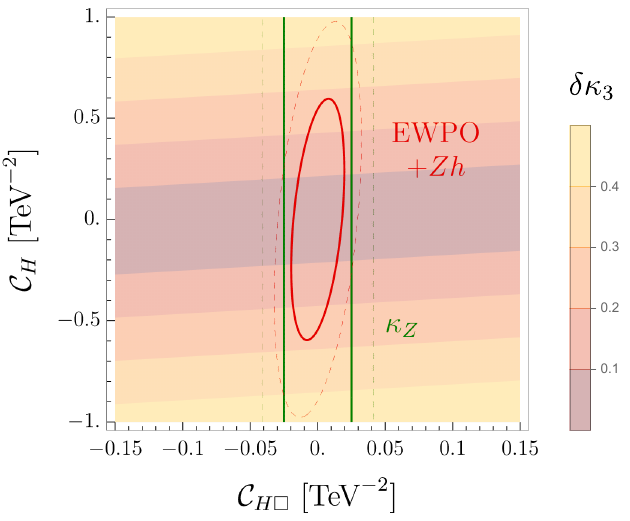}
    \caption{In red: allowed regions at 1$\sigma$ (solid line) and 2$\sigma$ (dashed line) confidence level from EWPOs and $e^+e^-\to Zh$ combined. In green: allowed regions for $\kappa_Z$.}
    \label{fig:kappacontours}
\end{figure}

When discussing generic BSM effects from heavy New Physics (NP) at and around the electroweak scale, it is useful to parametrise new interactions in the SM Effective Field Theory (SMEFT).
We normalise the Lagrangian as 
\begin{align}
    \mathcal{L}_{\rm SMEFT} = \mathcal{L}_{\rm SM} + \frac{1}{\Lambda^2}\sum_i \mathcal{C}_i \mathcal{O}_i \,,
\end{align}
where $\mathcal{C}_i$ indicates the Wilson coefficient, and $\mathcal{O}_i$ the effective operator.
The Higgs trilinear gets modified in SMEFT as follows:
\begin{align}
    \kappa_3 = 1 -  \frac{2 v^4}{\Lambda^2 m_h^2} \mathcal{C}_H + \frac{3v^2}{\Lambda^2}\left(\cC_{H\Box} - \frac{1}{4}\cC_{HD}\right) \,,
\end{align}
where
\begin{align}
    \nonumber
    \mathcal{O}_H &= (H^\dagger H)^3 \,, \\
    \nonumber
    \mathcal{O}_{H\Box} &= (H^\dagger H) \Box (H^\dagger H) \,, \\
    \mathcal{O}_{HD} &= |H^\dagger D_\mu H|^2 \,,
\end{align}
with $\Box \equiv \partial_\mu \partial^\mu$.
All three operators enter the $e^+e^-\to Zh$ cross-section, albeit not at the same perturbative order.
It is therefore clear that, even in the absence of other NP effects, it is not possible to extract $\kappa_3$ from the Higgstrahlung process alone, independently from the precision of the measurement.
Even considering the two planned FCC-ee runs (at 240 and 365\,GeV), and making use of the different energy dependence on the SMEFT coefficients, would not be conclusive, as one unconstrained direction will always remain.
Of course, the problem can be lifted by considering more observables, specifically electroweak precision observables (EWPOs).
Indeed, $\cC_{HD}$ is well known to be tightly constrained by tests of the $T$-parameter, which makes its contribution in $\kappa_3$ irrelevant.
The EW fit is also sensitive to $\cC_{H\Box}$ through loop effects, as discussed e.g. in Refs.~\cite{Barbieri:2007bh,Elias-Miro:2013eta,Contino:2013kra,Allwicher:2023shc,Maura:2024zxz}.
At FCC-ee, the integrated luminosity is higher for lower energies, resulting in a significantly larger dataset at the $Z$-pole than at the $Zh$ threshold.
In this context, EWPOs (loop-level) can have similar if not better sensitivity to $\cC_{H\Box}$ than $e^+e^- \to Zh$ (tree-level).
On top of that, the operator $\cC_{H\Box}$ leads to a tree-level, universal modification of all Higgs couplings.
In particular, the $hZZ$ coupling modifier $\kappa_Z$ is expected to be constrained at the per-mil level at FCC-ee~\cite{FCC:2025lpp}, leading to a constraint in the few-TeV range on $\cC_{H\Box}$ \footnote{In the EFT, the constrained quantity is always $\mathcal{C}/\Lambda^2$. When we report constraints on Wilson coefficients in TeV these should be interpreted as the bound on $\Lambda$ for $\mathcal{C}=1$, i.e. if for fixed $\Lambda$ we find $\mathcal{C} \in \{a,b\}$, then bound on the NP scale reads $\Lambda_{\rm NP} > \Lambda \times \text{min} (1/\sqrt{|a|},1/\sqrt{|b|})$}~\cite{Allwicher:2023shc}.
To illustrate these effects, we show in Fig.~\ref{fig:kappacontours} the expected allowed regions from FCC-ee in the $\cC_{H\Box}$--$\cC_H$ plane, superimposed with contours of $\delta\kappa_3$.
From that it is clear that the effect of $\cC_{H\Box}$ is also marginal, and will imply in the rest of this paper that $\delta\kappa_3\, \propto \,\cC_H$, such that a bound on $\cC_H$ can be automatically translated into a constraint on the Higgs self-coupling.
This is the approach followed in several works in the recent literature, among which Refs.~\cite{terHoeve:2025omu,Maura:2025rpe} are our main references for comparisons.

Assuming as a reference NP in $\mathcal{C}_H$ only, and a precision of $0.3\%$ in the measurement of the $Zh$ cross-section at FCC-ee (240\,GeV)~\cite{FCC:2025lpp}, one finds that the expected precision on $\kappa_3$ is
\begin{align}
    |\delta\kappa_3| \lesssim 0.17 
\end{align}
at the $1\sigma$ confidence level. We want to stress that this number differs from the 28\% quoted in the FCC Feasibility Study Report \cite{FCC:2025lpp,deBlas:2025gyz} for the sensitivity of FCC alone. This difference is due to the fact that we consider a modification of the trilinear only, while 28\% refers to the result of a global fit including all Higgs observables and coupling modifiers. Assuming an uncertainty on $\kappa_Z$ of 0.1\% in the $Zh$ cross-section one recovers the quoted FCC-ee sensitivity. Combining the FCC-ee sensitivity with the expectations from HL-LHC then yields a sensitivity around 18\%. In this work, we want to check how this determination of $\kappa_3$ can be affected when other NP effects are present and can also distort the Higgstrahlung process.

\section{Higgsstrahlung at $e^+e^-$ colliders}

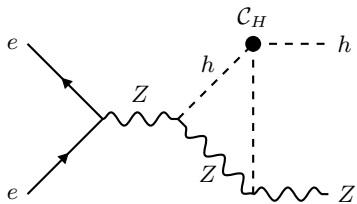
\begin{figure*}
        \centering
        \begin{tikzpicture}[thick,>=stealth]
            \node[wc] at (1,1) {};
            \node[above] at (1,1.1) {$\mathcal{C}_H$};
            \draw[midarrow] (-1,0) -- (-2,1) node[left] {$e$};
            \draw[midarrow] (-2,-1) node[left] {$e$} -- (-1,0);
            \draw[vector] (-1,0) -- (0,0);
            \node[above] at (-0.5,0.1) {$Z$};
            \draw[vector] (1,-1) -- (0,0);
            \draw[dashed] (0,0) -- (1,1);
            \draw[dashed] (1,1) -- (1,-1);
            \draw[dashed] (1,1) -- (2,1) node[right] {$h$};
            \draw[vector] (1,-1) -- (2,-1) node[right] {$Z$};
            \node[above] at (0.4,0.5) {$h$};
            \node[below] at (0.4,-0.5) {$Z$};
        \end{tikzpicture}
    \caption{Contribution of the Higgs trilinear coupling to the $Zh$ cross-section.}
    \label{fig:FeynDiagramsH}
\end{figure*}

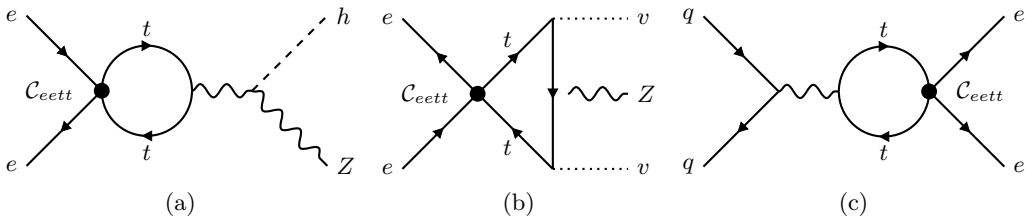
\begin{figure*}
    \begin{tabular}{ccc}
        \begin{tikzpicture}[thick,>=stealth]
            \draw[midarrow] (-2,1) node[left] {$e$} -- (-1,0);
            \draw[midarrow] (-1,0) -- (-2,-1) node[left] {$e$};
            \node[wc] at (-1,0) {};
            \node[left=7pt] at (-1,0) {$\cC_{eett}$};
            \draw[midarrow] (-1,0) arc (180:0:0.6);
            \node[above] at (-0.4,0.6) {$t$};
            \draw[midarrow] (0.2,0) arc (0:-180:0.6);
            \node[below] at (-0.4,-0.6) {$t$};
            \draw[vector] (0.2,0) -- (1,0);
            \draw[vector] (1,0) -- (2,-1) node[right] {$Z$};
            \draw[dashed] (1,0) -- (2,1) node[right] {$h$};
        \end{tikzpicture}
         &
        \begin{tikzpicture}[thick,>=stealth]
            \node[wc] at (0,0) {};
            \node[left=7pt] at (0,0) {$\cC_{eett}$};
            \draw[midarrow] (0,0) -- (-1,1) node[left] {$e$};
            \draw[midarrow] (-1,-1) node[left] {$e$} -- (0,0);
            \draw[midarrow] (1,-1) -- (0,0);
            \draw[midarrow] (0,0) -- (1,1);
            \draw[midarrow] (1,1) -- (1,-1);
            \draw[dotted] (1,1) -- (2,1) node[right] {$v$};
            \draw[dotted] (1,-1) -- (2,-1) node[right] {$v$};
            \node[above] at (0.4,0.5) {$t$};
            \node[below] at (0.4,-0.5) {$t$};
            \draw[vector] (1.2,0) -- (2,0) node[right] {$Z$};
        \end{tikzpicture}
         &
        \begin{tikzpicture}[thick,>=stealth]
            \draw[midarrow] (-2,1) node[left] {$q$} -- (-1,0);
            \draw[midarrow] (-1,0) -- (-2,-1) node[left] {$q$};
            \node[wc] at (1,0) {};
            \node[right=7pt] at (1,0) {$\cC_{eett}$};
            \draw[midarrow] (-0.2,0) arc (180:0:0.6);
            \node[above] at (0.4,0.6) {$t$};
            \draw[midarrow] (1,0) arc (0:-180:0.6);
            \node[below] at (0.4,-0.6) {$t$};
            \draw[vector] (-1,0) -- (-0.2,0);
            \draw[midarrow] (1,0) -- (2,-1) node[right] {$e$};
            \draw[midarrow] (1,0) -- (2,1) node[right] {$e$};
        \end{tikzpicture}
         \\
         (a) & (b) & (c)
    \end{tabular}
    \caption{Example one-loop SMEFT diagrams involving $eett$ operators relevant for some of the observables discussed in the main text. Figure~(a) exemplifies the contribution to Higgsstrahlung, (b) the RGE effects responsible for $Zee$-coupling modifications, and (c) NLO contributions in Drell--Yan production through valence quarks~($q$).}
    \label{fig:FeynDiagramseett}
\end{figure*}

Neglecting electron-mass effects and quadratic terms in dimension-six coefficients, the Higgstrahlung cross-section at an $e^+e^-$ collider has been recently fully computed in SMEFT at NLO, i.e. including loops involving Yukawa and gauge couplings~\cite{Asteriadis:2024xts}.
Following the same notation, the cross-section can be written as
\begin{align}
    \label{eq:eeZhxsec}
    \sigma_{e^+e^-\to Zh} = \sigma_{e^+e^-\to Zh}^{\rm NLO\,, SM} \left[1 + \sum_i \left(\Delta_i + \bar\Delta_i \log\frac{\mu^2}{\Lambda^2}\right)\mathcal{C}_i\right] \,,
\end{align}
where the $\mathcal{C}_i$ are in general combinations of SMEFT Wilson coefficients in the Warsaw basis~\cite{Grzadkowski:2010es}, $\Lambda$ is the scale of new physics (at which the coefficients are defined), and $\mu$ is the scale of the process ($\mu =\sqrt{s} = 240,365$\,GeV in our case). 
The numerical coefficients $\Delta_i$ come from tree-level or finite one-loop effects, while the logarithmically enhanced $\bar\Delta_i$ encode the divergent part of the loops, which is no other than the RGE effects (in the first leading-log approximation).
For the details of the computation and the numerical values, we refer to Ref.~\cite{Asteriadis:2024xts}.
In total, 10 Wilson coefficient combinations enter at tree-level, and an additional set of 29 at one-loop \footnote{This counting refers to the fully generic flavour scenario, i.e. with no specific flavour assumptions~\cite{Asteriadis:2024xts}. Among the operators entering at one-loop which are not considered in this ana\-lysis there are, in standard Warsaw notation~\cite{Grzadkowski:2010es}, $\cC_{H\ell}^{(1,3)}$, $\cC_{Hq}^{(1,3)}$, $\cC_{Hu,Hd}$ with different flavour indices (modifying the $Z$ couplings to fermions), the top dipoles $[\cC_{uB,uW}]_{33}$, the top Yukawa modifying operator $[\cC_{uH}]_{33}$, and several four-lepton and semileptonic operators, entering through diagrams analogous to Fig.~\ref{fig:FeynDiagramseett} (c). See Ref.~\cite{Asteriadis:2024xts} for the complete and detailed set.}.
Here, as mentioned in the introduction, we focus only on four-fermion interactions involving top quarks and electrons, i.e. the following 5 operators
\begin{align}
    \nonumber
    [\mathcal{O}_{lq}^{(1)}]_{1133} &= (\bar l_1 \gamma_\mu l_1)(\bar q_3 \gamma^\mu q_3) \,, \\
    \nonumber
    [\mathcal{O}_{lq}^{(3)}]_{1133} &= (\bar l_1 \gamma_\mu \sigma^I l_1)(\bar q_3 \gamma^\mu \sigma^I q_3) \,, \\
    \nonumber
    [\mathcal{O}_{qe}]_{3311} &= (\bar q_3 \gamma^\mu q_3)(\bar e_1 \gamma_\mu e_1) \,, \\
    \nonumber
    [\mathcal{O}_{lu}]_{1133} &= (\bar l_1 \gamma_\mu l_1)(\bar u_3 \gamma^\mu u_3) \,, \\
    [\mathcal{O}_{eu}]_{1133} &= (\bar e_1 \gamma_\mu e_1)(\bar u_3 \gamma^\mu u_3) \,,
    \label{eq:eettcoeff}
\end{align}
with the addition of $\cC_H$, responsible for the modification of the Higgs trilinear (see also the diagram in Fig.~\ref{fig:FeynDiagramsH}).

For the four-fermion operators the RG effects are usually dominant, with coefficients $\bar\Delta_i \approx 10^{-2}$, while $\mathcal{C}_H$ has a finite contribution only, with $\Delta_H = -7.21\times 10^{-3} (-9.98\times 10^{-4})$ at 240 (365)\,GeV (with unpolarised beams).

\section{Phenomenology of electron-top operators}\label{sec:eettpheno}

Besides modifying the $e^+e^-\to Zh$ cross-section, $eett$ operators are associated with a rich phenomenology, both at tree-level and at one-loop, which plays a crucial role in constraining them as we outline in the following (for a summary of the observables and inputs used see Table~\ref{tab:obs}). 
On top of that, the presence of the $b$ quark in the third-generation doublet opens the floor for discussing several complementary probes in flavour-violating and flavour-conserving transitions.
Example Feynman diagrams, displaying one-loop contributions of $eett$ operators to the different observables described in the following can be found in Fig.~\ref{fig:FeynDiagramseett}.

\begin{table}[]
    \centering
    \begin{tabular}{c|c}
        Obs. & Refs/comments \\ \hline
        Drell--Yan & \cite{Allwicher:2022mcg} (HL-LHC projection) \\
        EWPO & \cite{Breso-Pla:2021qoe} \\
        FCC-ee EWPO & \cite{FCC:2025lpp} \\
        $R_q$ (FCC-ee) & \cite{Greljo:2024ytg} 163, 240, and 365\,GeV \\
        $R_t$ (FCC-ee) & \cite{Greljo:2024ytg} 365\,GeV \\
        $R_e$ (FCC-ee) & \cite{Greljo:2024ytg} 163, 240, and 365\,GeV \\
        $e^+e^-\to Zh$ & \cite{FCC:2025lpp, Asteriadis:2024xts} 240 and 365\,GeV
    \end{tabular}
    \caption{Set of observables used in our analysis.}
    \label{tab:obs}
\end{table}

\subsection{Tree-level effects}

The most immediate tree-level effect mediated by the operators above is a modification of the $e^+e^-\to t\bar t$ cross-section. 
We include this observable through the ratio of cross-section at $\sqrt{s}=365$\,GeV
\begin{align}
    \nonumber
    R_t &= \frac{\sigma(e^+e^- \to t\bar t)}{\Sigma_{q=u,d,s,c,b\ } \sigma(e^+e^-\rightarrow q \bar{q})} \\ \nonumber
    &\simeq R_t^{\rm SM} + \left( 0.10 \ \cC_{lq}^{(3)} -0.17 \ \cC_{lq}^{(1)} -0.087 \ \cC_{lu} \right.\\ & \left.- 0.043\ \cC_{qe} - 0.095 \ \cC_{eu} \vphantom{\cC_{lq}^{(1)}} \right) \left(\frac{\textrm{1\,TeV}}{\Lambda}\right)^2\,,
    \label{eq:Rt}
\end{align}
where $R_t^{\rm SM} \simeq 0.11$, and we have kept only SM-NP interference terms, restricting to the five $eett$ operators. 
The ratio $R_t$ is expected to be measured at the 365\,GeV run of FCC-ee with $\mathcal{O}(10^{-4})$ precision~\cite{Greljo:2024ytg}.

Operators with the left-handed quark doublet, moreover, give rise to $eebb$ transitions, which can be probed through
\begin{enumerate}
    \item Drell--Yan at LHC, i.e. $pp\to ee$ with a $b\bar b$ pair in the initial state ($b\bar b \to ee$). This has been explored e.g. in Ref.~\cite{Allwicher:2022gkm}, yielding bounds in the TeV range.
    \item $b\bar b$ production at lepton colliders above the $Z$ pole, for which the FCC-ee prospects have been studied in detail in Ref.~\cite{Greljo:2024ytg}. 
\end{enumerate}
The observables used in the following for all fermion pair production from $e^+e^-$ are the ratios $R_a$, defined at different centre-of-mass energies, see Table~\ref{tab:obs},  as
\begin{align} \label{eq:Ra}
    R_a = \frac{\sigma(e^+e^-\rightarrow a \bar{a})}{\Sigma_{q=u,d,s,c,b} \ \sigma(e^+e^-\rightarrow q \bar{q})},
\end{align}
with $a = b,s,c,t$~\cite{Greljo:2024ytg}.

As a final remark, we want to note that $eett$ operators can also be probed by $t\bar t e^+e^-$ final states at the LHC. However, the constraints are not very strong compared to the others presented here~\cite{Grunwald:2025kot,Bellafronte:2025ubi}, and therefore not considered in our analysis.

\subsection{One-loop effects}\label{sec:oneloopeffects}

At one loop, operators with top quarks benefit from the enhancement of both the top Yukawa coupling and colour factors entering when closing quark loops.
These can be effectively accounted for by operator mixing induced by RGE effects, as has been pointed out e.g. in Ref.~\cite{Allwicher:2023shc}.
As an example, the $Z$ boson coupling to electrons is modified by the following terms~\footnote{Take for example $\mathcal{C}_{He}$. After electroweak symmetry breaking, $(H^\dagger i \overleftrightarrow D_\mu H)\supset -\frac{v^2}{2} Z_\mu$, such that there is a modification of the coupling of the $Z$ to right-handed electrons. In a similar fashion, the coupling to the left-handed electrons will be proportional to $[\mathcal{C}_{Hl}^{(1)}]_{11}+[\mathcal{C}_{Hl}^{(3)}]_{11}$.},
\begin{align}
    \nonumber
    \cL &\supset \frac{1}{\Lambda^2}[\cC_{He}]_{11} (H^\dagger i \overleftrightarrow D_\mu H)(\bar e_1 \gamma^\mu e_1) \\ \nonumber
    &+\frac{1}{\Lambda^2}[\cC_{Hl}^{(1)}]_{11} (H^\dagger i \overleftrightarrow D_\mu H)(\bar l_1 \gamma^\mu l_1) \\ 
    &+\frac{1}{\Lambda^2}[\cC_{Hl}^{(3)}]_{11} (H^\dagger i \overleftrightarrow D_\mu^I H)(\bar l_1 \gamma^\mu \sigma^I l_1)
    \,,
\end{align}
where $e$ refers to the right-handed $SU(2)$ singlet, and $l$ is the doublet in the usual Warsaw notation~\cite{Grzadkowski:2010es}.
Focussing for example on the right-handed coupling, the coefficient at the electroweak scale is approximately given by~\cite{Jenkins:2013wua}
\begin{align}\label{eq:HemZ}
    [\mathcal{C}_{He}]_{11} (m_Z) \approx \frac{y_t^2 N_c}{8\pi^2} \log\frac{m_Z}{\Lambda} \left([\cC_{qe}]_{3311}-[\cC_{eu}]_{1133}\right) \,,
\end{align}
where we have kept only the first leading log in solving the RGE, and neglected hypercharge-induced running.
Given the expected experimental precision at FCC-ee on the $Z$ pole, EWPOs provide another very relevant set of constraints.

Another noteworthy effect concerns Drell--Yan production at the LHC initiated by valence quarks.
This is a loop effect and qualitatively different from the one discussed in the previous paragraph.
Indeed, closing the top-quark loop in the $eett$ operators, one can get a contribution from $uu,dd \to ee$.
We include these effects in our analysis by including RGE effects from the high scale $\Lambda$ to the scale of each bin of the di-electron distribution, making use of {\tt HighPT}~\cite{Allwicher:2022mcg}.
There are also finite one-loop terms in SMEFT, which have been recently presented in Ref.~\cite{Bellafronte:2025ubi}.
We did not include the latter in our analysis, since we found that the RGE effects alone yield similar constraints as the one shown in Ref.~\cite{Bellafronte:2025ubi}, and were easier to take into account within the {\tt HighPT} framework. For a quantitative comparison between the two different effects we refer to the appendix.

\subsection{UV assumptions and additional correlated interactions} \label{sec:UVandRelatedStructures}

In this section, we comment briefly on the UV assumptions that are implicit in a scenario with NP in $eett$ operators, and their phenomenological implications.
A reasonable starting point is to impose on the SMEFT an appropriate flavour symmetry.
On the quark side, purely third-generation indices are compatible with a $U(2)$ symmetry acting on $q$ and $u$ fields, with the extra assumption of third-generation dominance, as for example studied in detail in Ref.~\cite{Allwicher:2023shc}. 
On the lepton side, selecting electron coefficients may be achieved by imposing electron number conservation, i.e. a $U(1)_{e}$ symmetry (acting equally on left- and right-handed first generation leptons).
Another option would be to assume a $U(2)$ symmetry on the lepton side as well, i.e. going towards the full $U(2)^5$ scenario.
This would have as a consequence that operators with electrons and muons always come together, which is why a purely $U(1)_e$ scenario is likely to be more conservative.

Proceeding with the top- and electrophilic hypothesis, other operators in the EFT may be written down.
Although these do not contribute to the Higgsstrahlung process (in the limit of $m_e=0$), they may be phenomenologically relevant for the other observables discussed above.
In the following we briefly go through them and discuss their effects.

\subsubsection{Scalar and tensor operators}
The scalar and tensor operators complying with our assumptions are
\begin{align}
    \nonumber
    [\mathcal{O}_{lequ}^{(1)}]_{1133} &= (\bar l_1 e_1) (\bar q_3 u_3) \\
    [\mathcal{O}_{lequ}^{(3)}]_{1133} &= (\bar l_1 \sigma_{\mu\nu} e_1) (\bar q_3 \sigma^{\mu\nu} u_3)
    \label{eq:scalarOps}
\end{align}
These have top-quark Yukawa enhanced running effects into both the electron Yukawa and electron dipole operators (the latter one being two-loop), so one gets a constraint from the electron $g-2$ \cite{Crivellin:2020mjs,Davoudiasl:2023huk,Allwicher:2025mmc}.
However, neither of these operators do enter the $Zh$ cross-section within our working assumption (interference terms only, $m_e=0$, one-loop), and they do not have an effect in EWPOs or Drell--Yan either. 
The presence of a right-handed top also prevents them from contributing to any relevant $B$-physics observable.
The only place where they may play a role is $e^+e^-\to t\bar t$ transitions.

\subsubsection{Four-lepton and four-quark operators}

Under the same assumptions, four-lepton and four-quark operators may be generated as well.
In particular, the following structures would be allowed
\begin{align}
    \nonumber
    [\mathcal{O}_{ee}]_{1111} &= (\bar e_1 \gamma_\mu e_1)(\bar e_1 \gamma^\mu e_1) \,, \\
    \nonumber
    [\mathcal{O}_{le}]_{1111} &= (\bar l_1 \gamma_\mu l_1)(\bar e_1 \gamma^\mu e_1) \,, \\
    \nonumber
    [\mathcal{O}_{ll}]_{1111} &= (\bar l_1 \gamma_\mu l_1)(\bar l_1 \gamma^\mu l_1) \,, \\
    \nonumber
    [\mathcal{O}_{qq}^{(1)}]_{3333} &= (\bar q_3 \gamma_\mu q_3)(\bar q_3 \gamma^\mu q_3) \,, \\
    \nonumber
    [\mathcal{O}_{qq}^{(3)}]_{3333} &= (\bar q_3 \gamma_\mu \sigma^I q_3)(\bar q_3 \gamma^\mu \sigma^I q_3) \,, \\
    \nonumber
    [\mathcal{O}_{uu}]_{3333} &= (\bar u_3 \gamma_\mu u_3)(\bar u_3 \gamma^\mu u_3) \,, \\
    \nonumber
    [\mathcal{O}_{qu}^{(1)}]_{3333} &= (\bar q_3 \gamma_\mu q_3)(\bar u_3 \gamma^\mu u_3) \,, \\
    [\mathcal{O}_{qu}^{(8)}]_{3333} &= (\bar q_3 \gamma_\mu T^A q_3)(\bar u_3 \gamma^\mu T^A u_3) \,.
    \label{eq:4l4qOps}
\end{align}
The four-electron operators will be tightly constrained by $e^+e^-\to e^+e^-$ (through $R_e$, defined the same way as Eq.~\ref{eq:Ra}) scattering at FCC above the $Z$ resonance~\cite{Greljo:2024ytg}, yielding bounds in the 50\,TeV range.
The four-quark operators, on the other hand, will mainly contribute to EWPOs through top-quark Yukawa-enhanced RGE effects.
The largest effects will therefore arise in modifications of the $Zbb$ vertex, in analogy to the electron case discussed above.
Typical expected bounds here lie in the 5\,TeV ballpark from EWPOs at FCC-ee, while current direct constraints from $b\bar b$ and $t\bar t$ final states at the LHC are around the TeV scale~\cite{Ethier:2021bye,Celada:2024mcf}.

\subsection{Flavour constraints}

Precision flavour observables, $B$-decays especially, can provide additional constraints.
With the given choice of operators, flavour-violating NP effects can arise via i) misalignment in the quark doublet $q_3$, or ii) loop effects, i.e. when the flavour violation is mediated by the SM Yukawas.
The alignment issue always arises in SMEFT due to the fact that the weak interaction eigenbasis and the mass eigenbasis do not coincide.
In other words, one always needs to specify a basis for the Yukawa matrices.
In our somewhat simplified context, this boils down to defining the composition of the third-generation quark doublet (see Ref.~\cite{Allwicher:2023shc} for an extensive discussion).
A conservative choice consists in assuming down-quark alignment, i.e. the limit
\begin{align}
    q_3 = \begin{pmatrix}
        \sum_i V_{3i}^* u_L^i \\
        b_L
    \end{pmatrix} \,,
\end{align}
where $V$ is the CKM matrix.
In this case, the only tree-level flavour-violating effects arise in the up-quark sector, and are relevant only for four-quark operators leading to neutral $D$-meson mixing.
Current bounds lie in the TeV range and below~\cite{Allwicher:2023shc}, with future prospects (including FCC-ee) improving the reach by a factor of 2$\div$3~\cite{Allwicher:2025bub}.

Loop effects, on the other hand, can be taken into account again through the RGE, the flavour violation now arising from the SM Yukawas in the loops.
Given our setup, the main effects would be in $b\to s ee$ transitions, e.g. $B\to K^{(*)}ee$ decays.
The discussion in Ref.~\cite{Bordone:2025cde} shows that the projected precision at FCC-ee can bring the sensitivity close to the one from EWPOs for $\sim 20\%$ down-alignment.
In our case, the amount of misalignment arising from loops is, parametrically \footnote{The parameter $\epsilon_F$ denotes the amount of down-alignment in the third-generation left-handed quark doublet. The limit $\epsilon_F=0$ is full down alignment, and $\epsilon_F=1$ is up-alignment. The analysis in Ref.~\cite{Bordone:2025cde} shows that flavour and EWPOs compete for $\epsilon_F \gtrsim 0.3$.},
\begin{align}
    \epsilon_F \sim \frac{1}{4\pi}y_t\sqrt{V_{ts}} \simeq 0.015 \,,
\end{align}
from which we expect flavour transitions to have a minor impact.

\section{SMEFT fits}\label{sec:SMEFT}

Our final goal is to assess the impact of $eett$ interactions on the determination of $\kappa_3$.
Some bounds on these interactions have been already presented in Ref.~\cite{Bellafronte:2025ubi}.
The most striking feature of that analysis is the presence of a somewhat flat direction appearing in the global fit, i.e. in the marginalised bound on the Wilson coefficients.
These are around the TeV mark, and a factor of 4$\div$5 worse than the individual fits, suggesting a relatively unconstrained direction.
In the context of the extraction of the Higgs trilinear, bounds in the TeV range allow for large NP effects in the $Zh$ cross-section (at the few-percent level, cf. Eq.~\ref{eq:eeZhxsec}).
We find that the flat direction in Ref.~\cite{Bellafronte:2025ubi} seemingly appears when including all 5 $eett$ coefficients, implying that one linear combination of coefficients is poorly constrained.
However, this degeneracy can be resolved by adding independent observables in the analysis, which in our case are the $R_q$ ratios above the $Z$-pole discussed in the previous section.
The addition of these observables represents the most important difference of this analysis with respect to Ref.~\cite{Bellafronte:2025ubi}.
In the following, we first discuss the details about the constraints on $eett$ interactions, and then move to the effects in the measurement of $\kappa_3$.

\subsection{Constraints on $eett$ operators}

\begin{figure} [h]
    \begin{subfigure}[c]{0.49\textwidth}
        \centering
        \includegraphics[width=\linewidth]{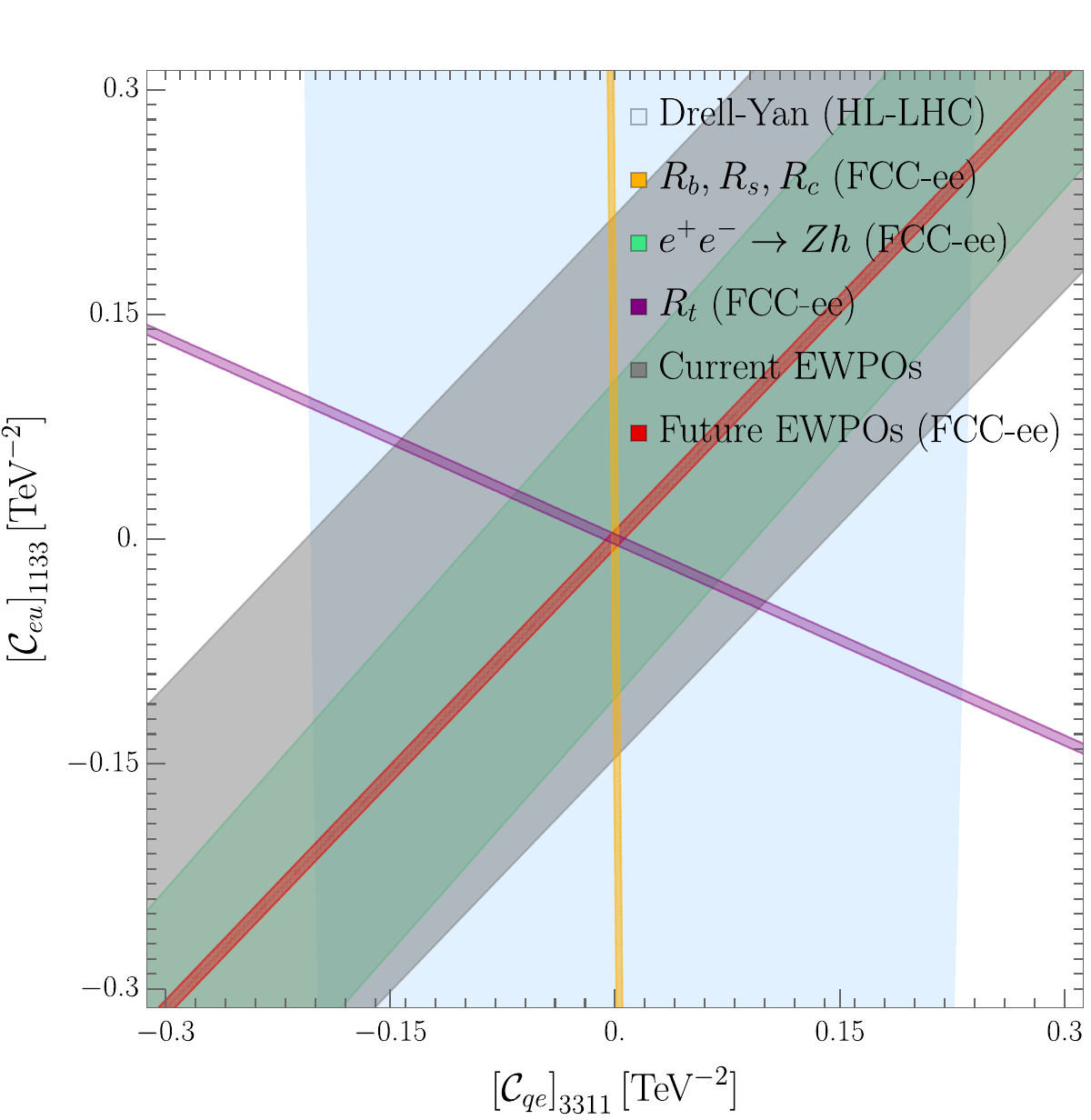}
    \end{subfigure}
    \begin{subfigure}[c]{0.49\textwidth}
        \centering
        \includegraphics[width=\linewidth]{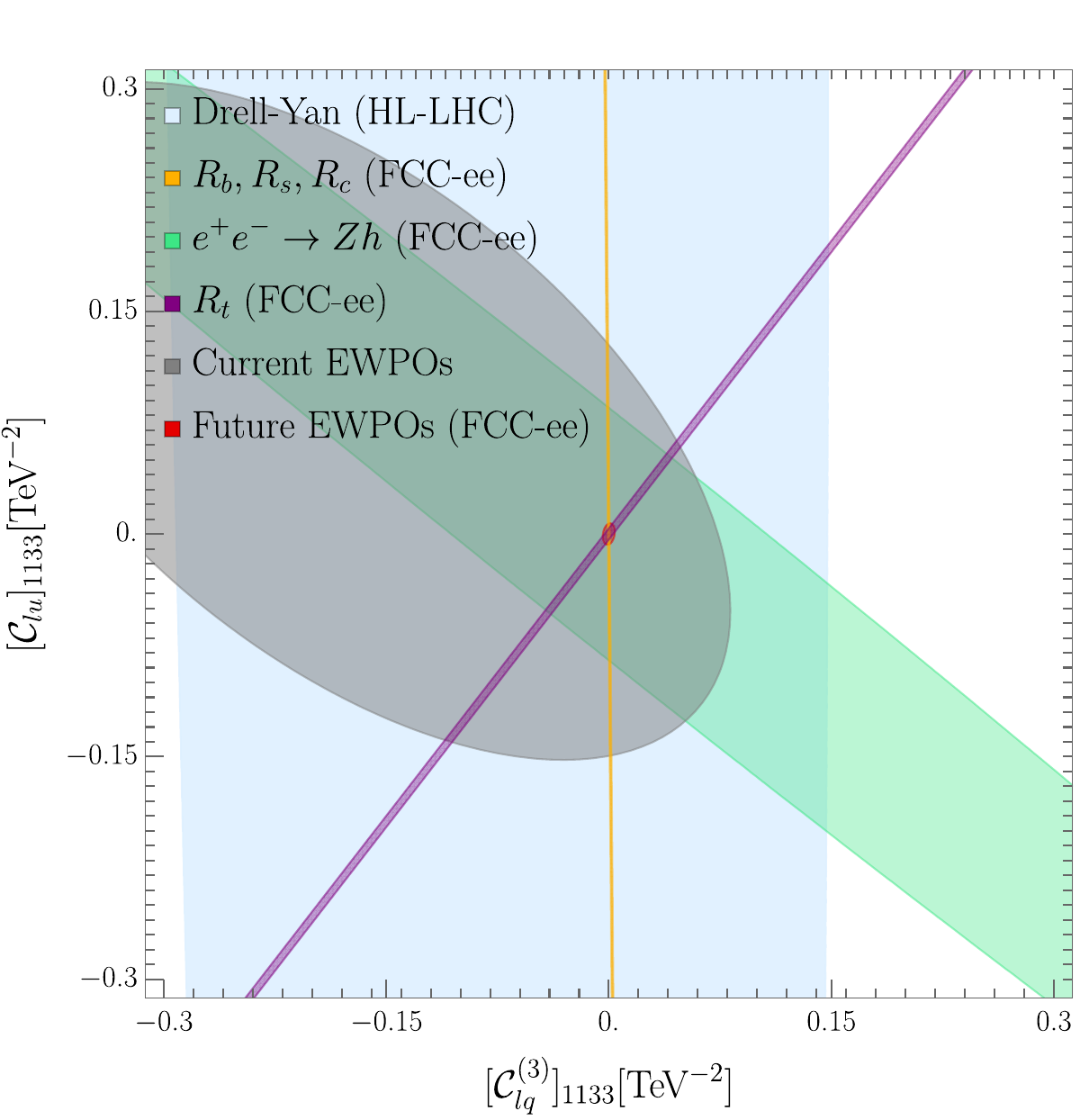}
    \end{subfigure}
    \caption{The 2$\sigma$ C.L. allowed regions from different processes for two of the coefficients. The left plot shows the coefficients with right handed leptons, $[C_{eu}]_{1133}$ and $[C_{qe}]_{1133}$ and on the right side coefficients with left handed leptons, $[C_{lu}]_{1133}$ and $[C_{lq}^{(3)}]_{3311}$. Respectively all other SMEFT coefficients are set to zero.}
    \label{fig:2CRegionPlot}
\end{figure}

In order to get a quantitative idea about the interplay of the different effects discussed in the previous section, we start by restricting the possible parameter space of two effective coefficients at a time, and study different combinations while setting all other coefficients to zero.
To do so, we assume that all coefficients are defined at a scale $\Lambda=1$\,TeV, and take into account RGE effects (in the first leading-log approximation) to the respective scale of each observable by means of {\tt DsixTools}~\cite{Fuentes-Martin:2020zaz}.
The results for $\cC_{qe}$--$\cC_{eu}$ and $\cC_{lq}^{(3)}$--$\cC_{lu}$ respectively are shown in Fig.~\ref{fig:2CRegionPlot}, where we display the allowed regions for all relevant observables. For a better comparison to other literature we restrict the coefficients at 2$\sigma$.

A few comments are in order:
\begin{itemize}
    \item[(i)] In the left plot of Fig.~\ref{fig:2CRegionPlot}, the regions allowed by $e^+e^-\to Zh$ and electroweak precision are roughly aligned. This is easily explained by the fact that both are in fact sensitive to the same operator at tree-level (the one modifying the coupling of the $Z$ to right-handed electrons).
    Under the assumptions that 1) top-Yukawa running effects dominate and 2) in the electroweak fit the right-handed and left-handed electron couplings can always be disentangled, this statement is easily generalised to any other combination of WCs.
    The small observable differences (leading to the misalignment of the two regions) are due to the finite terms computed in Ref.~\cite{Asteriadis:2024xts} \footnote{Notice also that here we are not including the analogous finite terms in the EWPOs.}.
    The right plot in Fig.~\ref{fig:2CRegionPlot} highlights the only interesting exception. Here there is no flat direction in the EW observables. This can be traced back to the fact that $\cC_{lq}^{3}$ will always (and alone) receive the additional constraint coming from $W$ decays ($W\to e\nu$ in this case). This point will also be relevant below, when considering marginalised fits over the full set of $eett$ operators.
    Finally, another important observation is that for the relevant region of parameter space the energy-dependent effects in Higgsstrahlung usually invoked to disentangle correlated effects are of no consequence in this case, as they would be of help only for large values of the WCs. 
    \item[(ii)] The constraints from $ee\to tt$ (through the ratio $R_t$) tend to be orthogonal to the ones from EWPOs. 
    This is readily seen in the case of right-handed electrons. 
    In EWPOs, the coupling of the $Z$ boson to $e$ is modified by $[\cC_{He}]_{11}$ (see discussion in Section \ref{sec:oneloopeffects}), whose contribution from $eett$ operators comes at one-loop and can be approximated as in Eq.~(\ref{eq:HemZ}).
    A quick look at the expression for $R_t$ in 
    Eq.~(\ref{eq:Rt}) then clearly shows how the two sets of observables are sensitive to orthogonal combinations of $\cC_{qe}$ and $\cC_{eu}$.
    Since it is known that the EW fit is able to constrain $\cC_{Hl}^{(1,3)}$ separately as well, this statement can be readily generalised.
    Indeed, one has that
    \begin{align}
        \left.[\dot\cC_{Hl}^{(1)}]_{11}\right|_{y_t} \propto \,[\cC_{lq}^{(1)}]_{1133} - [\cC_{lu}]_{1133} \,,
    \end{align}
    while the triplet operator $[\cC_{Hl}^{(3)}]_{11}$ is again constrained independently (see comment (i)).
    \item[(iii)] A very important constraint comes from $ee\to bb$ at energies above the $Z$ resonance, mainly through the ratio $R_b$ \footnote{$R_s$ and $R_c$ also depend on $ee\to b\bar b$ through the normalisation of the ratios (see Eq.~(\ref{eq:Ra})). However this contribution is suppressed by $R_b^{\rm SM} \simeq 0.15$.}. As argued in Ref.~\cite{Greljo:2024ytg}, this can be competitive, and, as we clearly see  complementary also, to constraints from $Z$-pole observables. This is especially true for operators which give $ee\to bb$ at tree-level, while the corresponding $Z\to ee$ modification is a one-loop effect. 
    Here the higher luminosity compensates the loop suppression.
    \item[(iv)] Comparing the expected Drell--Yan region at the end of HL-LHC with e.g. the corresponding one in Ref.~\cite{Bellafronte:2025ubi}, we notice that, on the $\cC_{qe}$ side, the constraint from tree-level $bb\to ee$ is much tighter than the one from one-loop $uu,dd\to ee$. On the other hand, even though it cannot be appreciated on the plot due to the overall scale, for $\cC_{eu}$ the size of the finite terms and the RGE effects are comparable.
    See also the appendix for a more detailed comparison.
    \item[(v)] In both plots, the main constraints come from EWPOs, $ee\to bb$, and $ee\to t\bar t$. These are enough to remove any flat directions and restrict the parameter space to a very small region.
    \item[(vi)] $e^+e^-\to Zh$ is never a competitive constraint for these operators, which seems to already indicate that the extraction of the Higgs trilinear might be affected only moderately by the presence of $eett$ interactions.
\end{itemize}
All of the above, and especially the last two statements, deserve some further and more general investigation, in order to check if they hold in a more global analysis with a larger number of operators.

\begin{figure*}
    \centering
    \includegraphics[width=0.9\linewidth]{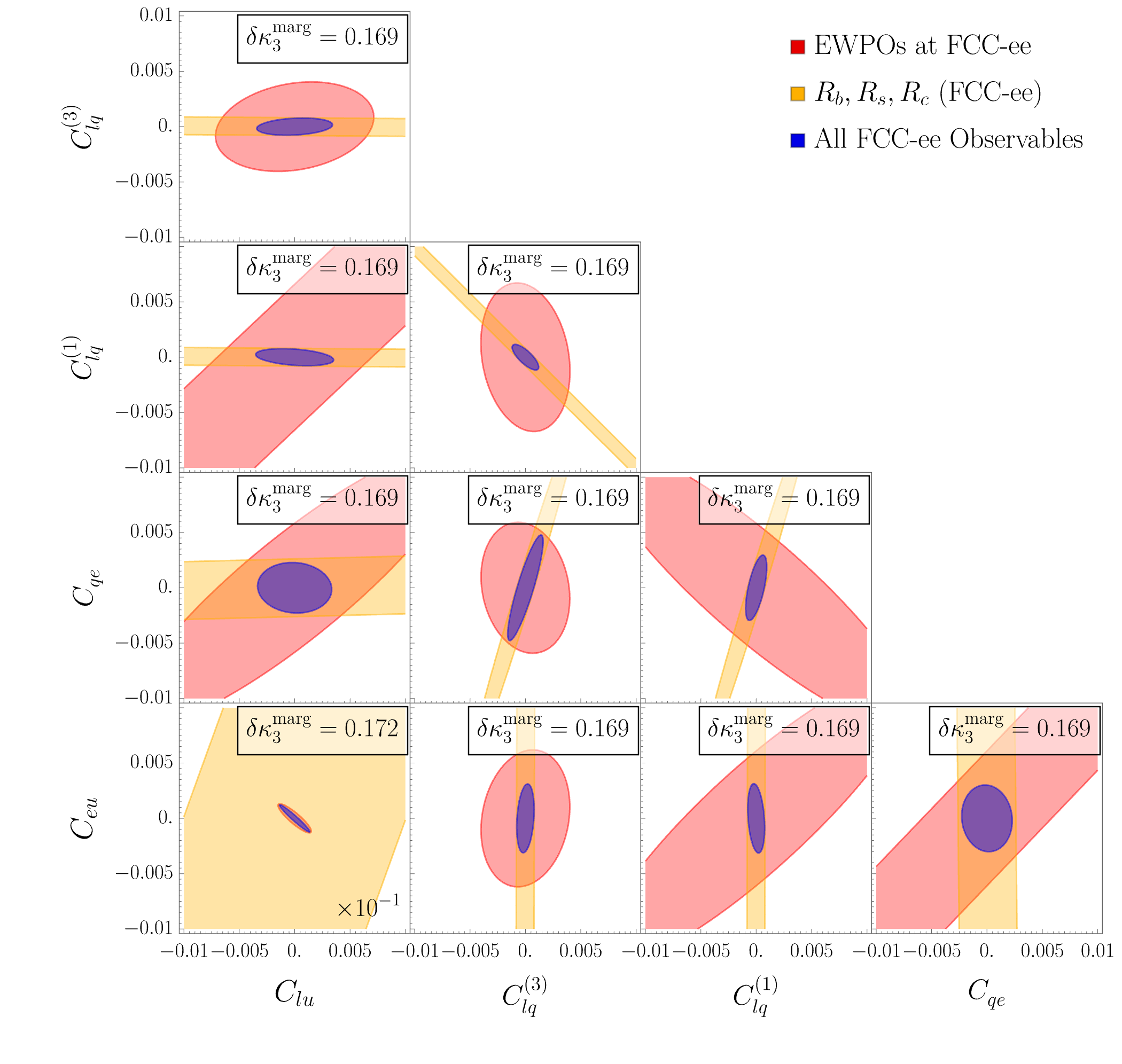}
    \caption{2$\sigma$ C.L. allowed regions from different observables for all combinations of two $eett$ coefficients.
    All the other SMEFT coefficients for each plot are set to zero . The coefficents are given in 1/TeV$^2$. The quoted values of $\delta\kappa_3$ (at 1$\sigma$) refer to the sensitivity obtained when marginalising over the given pair of coefficients.
    }
    \label{fig:plotmatrix2Coeff}
\end{figure*}

To do so, we show all possible combinations of two coefficients in Fig.~\ref{fig:plotmatrix2Coeff}.
For clarity, we have only included FCC-ee observables, showing EWPOs and $ee\to bb$ separately, and the combination of all observables, which includes $Zh$ production and, most importantly, $R_t$. Drell--Yan effects, on the other hand, are never competitive and therefore not shown (cf. again Figure \ref{fig:2CRegionPlot}).
The results show tight constraints for all combinations, and highlight again the importance of fermion pair production above the pole for this scenario.
The only exception is the $\cC_{lu}$--$\cC_{eu}$ pair, which gives no tree-level contribution in $ee\to bb$.
The allowed band for the EWPOs is then cut off by $t\bar t$ production at 365\,GeV.

\begin{figure*}
    \centering
    \includegraphics[width=\linewidth]{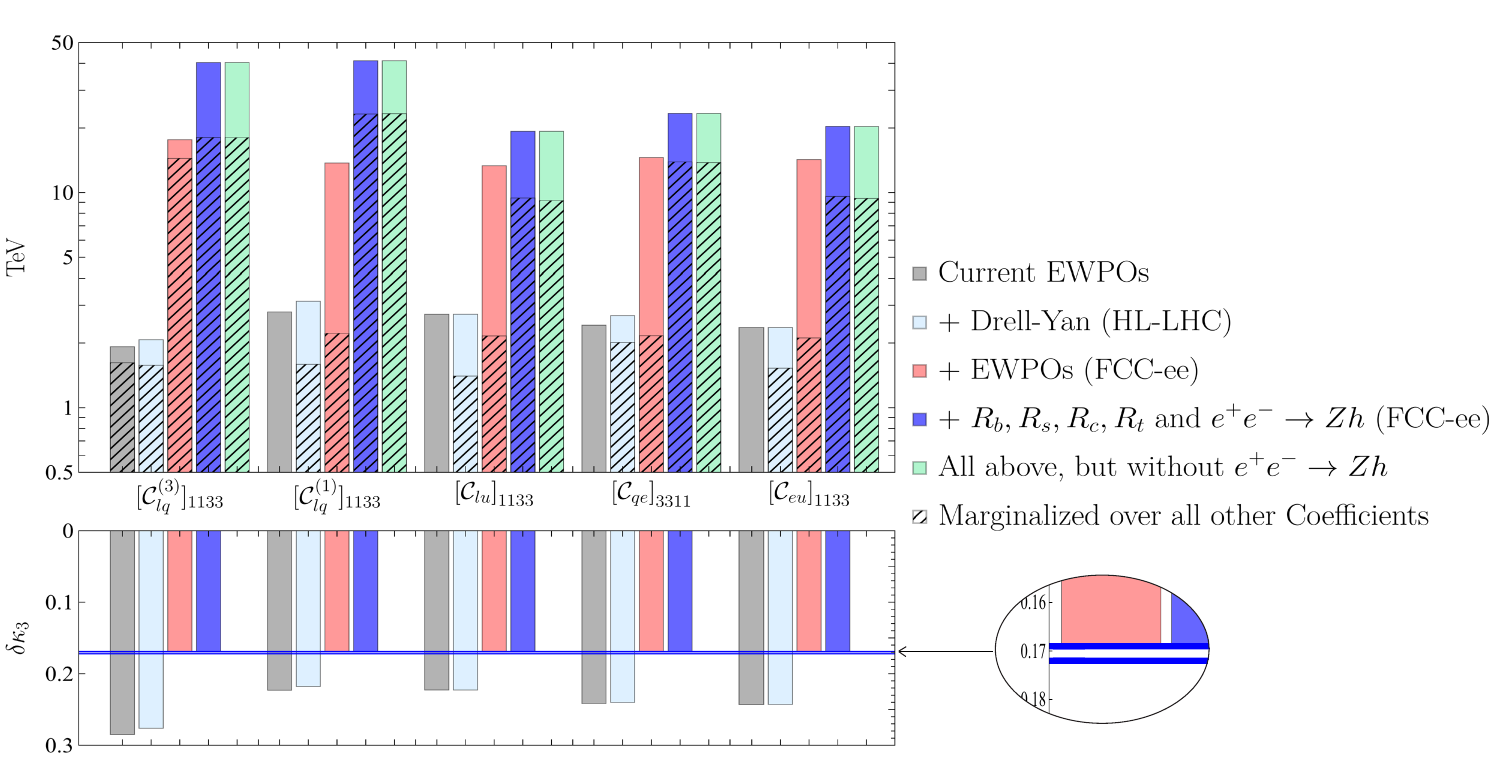}
    \caption{\textit{Upper panel:} Bounds on the different $eett$ coefficients. The solid colors represent $2\sigma$-bounds coming from individual coefficient fits. The hatched bars are the respective $2\sigma$-bounds obtained through marginalizing over all other $eett$ coefficients. For a detailed explanation of the observables, see main text. \textit{Lower panel:} 1$\sigma$ bounds on $\delta \kappa_3$ marginalised over the respective $eett$ coefficient. The observable $e^+e^- \rightarrow Zh$ is always included in order to get the bounds on $\delta \kappa_3$. The  nearly indistinguishable two blue lines represent the $1\sigma$ bounds $\delta \kappa_3^{\text{sep}}= 0.169$ and $\delta \kappa_3^{\text{marg}}= 0.172$, the first calculated with no $eett$ coefficient turned on and the second marginalised over all $eett$ coefficients, with all observables.}
    \label{fig:BoundsMargSep}
\end{figure*}

We furthermore performed a global fit with all five coefficients activated, and show the results in Fig.~\ref{fig:BoundsMargSep} and Tab.~\ref{tab:eettbounds}.
As expected, marginalising over all other coefficients leads to weaker bounds than individual WC fits (hatched against filled bars in the chart). However, we do not observe the large drop in sensitivity that can be seen in Ref.~\cite{Bellafronte:2025ubi}.
The drop itself is only observed as soon as all coefficients except $\cC_{lq}^{(3)}$ are activated\footnote{As anticipated above, the case of $\cC_{lq}^{(3)}$ is special due to the additional constraint from $W$ decays, which allows to always isolate it.}.
However, the approximately flat direction is lifted with the inclusion of $b\bar b$ production.
Another important information coming from this analysis is again the fact that $Zh$ production does not play a crucial role in this fit, which is easily understood given the fact that the constraints tend to align with the ones from EWPOs, but the expected experimental precision is lower for Higgstrahlung.

Referring back to Section~\ref{sec:UVandRelatedStructures}, we also checked the restriction of the parameter space turning on the scalar, Eq.~(\ref{eq:scalarOps}), as well as the four-lepton and four-quark operators, Eq.~(\ref{eq:4l4qOps}). In the same way as in Fig.~\ref{fig:BoundsMargSep}, with the addition of the ratio $R_e$, we calculate the marginalized bounds and report them alongside the previous ones in Tab.~\ref{tab:eettbounds}. As expected from the larger parameter space the bounds get weaker by roughly a factor 3$\div$4. Nevertheless, they show no flat direction, and the impact on the extraction of the Higgs trilinear, as we will show, is also marginal.
In the following we will go back to focussing only on the five $eett$-operators.

Before moving on to the discussion on the Higgs self-coupling, we want to briefly address the choice of the scale for the running, which is set to be $\Lambda=1$\,TeV.
Looking at the constraints \textit{a posteriori}, it would seem that a scale in the 10\,TeV range would be more appropriate.
However we would like to point out that i) for individual bounds, the phenomenology is completely dominated by tree-level effects, i.e. the logarithms have little impact, ii) for our marginalised fits, where loop-induced effects are relevant for the removal of flat direction, we checked that for a scale $\Lambda=10$\,TeV the bound change at most by 20$\div$30\%, and always get stronger (our conclusions are therefore conservative), and finally iii) the choice of 1\,TeV allows for an easier comparison with the previous literature.

\subsection{Impact on the Higgs trilinear}

We now want to quantify the effects of $eett$ interactions for the Higgs trilinear.
Before proceeding, we want to stress that similar analyses have been performed in the recent literature, in particular we mention Ref.~\cite{Maura:2025rpe}, which studied the same question but from a more global perspective, i.e. considering the full SMEFT dimension-six basis, imposing different flavour-symmetry and -assumptions scenarios.
Our analysis on the other hand focusses on electron-top interactions only (as considered in Ref.~\cite{Bellafronte:2025ubi}).
Their findings are in line with what we find in this section, as we will discuss in more detail.
On the other hand, Ref.~\cite{terHoeve:2025omu} tackled the issue from the perspective of the bosonic sector of the SMEFT, which has the largest effects in Higgstrahlung. 
They also report similar findings in the top-centric flavour assumption for the SMEFT.
Both analyses make use of the NLO computation of the $Zh$ cross-section~\cite{Asteriadis:2024xts}, and take into account RGE effects in the same manner as in this work.
The latter also includes finite matching terms in the UV for some selected models involving scalar electroweak quadruplet states.
We include a discussion along similar lines, although for different BSM states, in the next section.

Assuming that $\kappa_3$ only receives contributions from $\cC_H$,
we find that the $Zh$ production, under the assumption of a SM-like signal, constrains the trilinear coupling modifier to~\footnote{All quoted values for $\delta\kappa_3$ are to be understood as defined at 1 TeV, as we directly use the constraints on $\cC_H$ at that scale.}
\begin{align}
    |\delta\kappa_3| \leq 16.9 \%
\end{align}
at $1\sigma$ C.L. (cf. Table~\ref{tab:deltakappalambda}). 
Notice how this appears to be much smaller than the projected sensitivity quoted in the FCC feasibility study, which was coming from a global fit to all Higgs couplings, and yielding a sensitivity of 28\%~\cite{FCC:2025lpp} \footnote{Our number is however accidentally very similar to the one obtained combining FCC-ee with HL-LHC ($\sim$18\%).}.
Next, we proceed to switch on one $eett$ operator at a time, and fit $\cC_H$ marginalising over the respective four-fermion operator.
The results are summarised in Table~\ref{tab:deltakappalambda}, showing no significant effect in the extraction of $\kappa_3$.
Marginalising over all $eett$ coefficients instead leads to an approximate relative 2\% (relative) worsening of $\delta\kappa_3$.
This is in line with the effective scales in Fig.~\ref{fig:BoundsMargSep}, where one should keep in mind that a $\sim 50\%$ deviation in the Higgs trilinear roughly corresponds to TeV-scale New Physics~\cite{Maura:2025rpe}.
Furthermore, the lower half of Fig.~\ref{fig:BoundsMargSep} shows the 1$\sigma$ bounds on $\delta\kappa_3$ for different observables, with always one $eett$ coefficient activated. To get a bound on $\kappa_3$, the measurement on $e^+e^-\rightarrow Zh$ is always included. We can observe a stabilization around the single-coefficient value (upper blue line) once the EWPOs at FCCee are included.
Figure~\ref{fig:BoundsMargSep}, which is in direct analogy with the one presented in Ref.~\cite{Bellafronte:2025ubi}, deserves some further commenting.
We observe a very similar behaviour of the individual vs. marginalised bounds (cf. especially the red bars, corresponding to the inclusion of FCC-ee EWPOs, but not the above-pole $R_q$-ratios, i.e. exactly as in Ref.~\cite{Bellafronte:2025ubi}).
The addition of $R_q$ changes the picture significantly, since the difference between marginalised and individual fits shrinks sizeably, and on top of that the effective scales probed are higher, reaching 50\,TeV for individual fits (in line with Ref.~\cite{Greljo:2024ytg}) and $\sim$20 TeV for marginalised fits~\footnote{The important feature of adding the $R_q$ ratios lies in the lifting of the flat direction. The scales effectively probed have a smaller impact on our results, as we find that even assuming inflated uncertainties by an order of magnitude the extraction of $\kappa_3$ is almost unaffected, changing from $\delta\kappa_3=0.172$ to $\delta\kappa_3=0.173$.}.
The individual bounds on the five $eett$-operators (see again Table~\ref{tab:eettbounds} and Fig.~\ref{fig:BoundsMargSep}) are also comparable with the ones obtained in Ref.~\cite{cornetgomez2025futurecolliderconstraintstopquark} for FCC-ee projections. 

Finally we notice how the green bars, obtained with the full set of observables but without the $Zh$ cross-section, are essentially identical to the blue ones.
This shows, on one hand, how Higgstrahlung does not play a role in constraining $eett$ operators, and, on the other hand, why we should not expect these operators to impact the extraction of the Higgs trilinear, 
as indeed we observe numerically.

Finally, allowing all top- and electrophilic interactions, as described in Section~\ref{sec:UVandRelatedStructures}, we arrive at the more conservative bound
$\delta\kappa_3 \lesssim$ 23\%.  This is comparable with the 25\% bound described in Ref.~\cite{Maura:2025rpe}, when assuming a $U(2)^5$ symmetry with a dominance of the 3rd generation.\footnote{This seems reasonable given the fact that EWPOs play a crucial role in this fit. In the third-generation dominance scenario the coupling to electrons is suppressed, but the precision on $Z\to \tau\tau$ observables is comparable. What is suppressed is the contribution from fermion pair production above the $Z$ resonance, which is likely a contributing factor in the (still mild) loss of sensitivity.} This drop can be traced back to the four lepton operators, mainly $\left[\mathcal{O}_{ee}\right]_{1111}$ in combination with one more four lepton operator, as we only include one observable ($R_e$) constraining these operators.
Nevertheless, the number remains small, showing again the already described stability in measuring $\kappa_3$, when opening up the considered setting.

\begin{table}[]
    \centering
    \renewcommand{\arraystretch}{1.2}
    \begin{tabular}{c|c|c|c}
         &  & Marg. & Marg. \\
        Coeff. & Indiv. & (all $eett$) & (all top- \& electrophil.) \\ \hline
        $\cC_{lq}^{(1)}$ & 40.3 & 18.0 & 3.7 \\
        $\cC_{lq}^{(3)}$ & 41.1 & 23.2 & 5.0 \\
        $\cC_{lu}^{}$ & 19.3 & 9.4 & 4.0 \\
        $\cC_{qe}^{}$ & 23.3 & 13.9 & 4.0 \\
        $\cC_{eu}^{}$ & 20.3 & 9.6 & 2.8 \\
    \end{tabular}
    \caption{Lower bounds on $eett$ coefficients (in TeV).}
    \label{tab:eettbounds}
\end{table}

\begin{table}[]
    \centering
    \renewcommand{\arraystretch}{1.2}
    \begin{tabular}{c|c}
         non-zero SMEFT-Coefficients (on top of $\cC_H$) &  $\delta\kappa_3$\\ \hline
         None & 0.169 \\
         One $eett$-coefficient & 0.169 \\
         All $eett$-coefficients & 0.172 \\
         All $eett$ (w/o $R_q$ above pole) & 0.176 \\
         All top- and electrophilic coefficients & 0.23 \\
         $U(2)^5$, third-gen. dominance~\cite{Maura:2025rpe} & 0.25
    \end{tabular}
    \caption{
    The $1\sigma$ C.L. bounds on the trilinear Higgs coupling modifier, turning on a different number of coefficients. The bounds are obtained marginalising over the operators listed in the left column, with all observables defined in Table~\ref{tab:obs}. The only contributing Wilson coefficient for $\kappa_3$ is assumed to be $\mathcal{C}_H$.
   }
    \label{tab:deltakappalambda}
\end{table}

\section{Simplified Models}\label{sec:models}

In this section, we move away from the pure SMEFT analysis presented so far, and go one step further in the UV, assuming the presence of new heavy BSM states.
The easiest way of generating new physics effects in the $eett$ operators without the four-lepton or four-quarks counterparts is 
to UV complete the SM with one more or more leptoquarks,
In this way, only semileptonic operators are generated at tree-level, while every other type of four-fermion operator, or even fermion-Higgs operators, are generated starting from one loop.
Following the notation of 
Ref.~\cite{Dorsner:2016wpm}, we choose to focus on two possibilities, the scalar states $S_1 \sim (\mathbf{3},\mathbf{1},-1/3)$ and $R_2 \sim (\mathbf{3},\mathbf{2},7/6)$, and assume couplings to first generation leptons and third generation quarks only.

\noindent
The relevant Lagrangians of the Leptoquarks are \footnote{The $S_1$ leptoquark is also a di-quark, with couplings both to left-handed and right-handed pairs of quarks ($qq^c$ and $du^c$ respectively). Here we ignore these couplings, since they generally give a contribution to proton decay. In a more complete UV setting, these may be forbidden e.g. by requiring baryon number conservation.}
\begin{align}
    -\mathcal{L}_{S_1} &= [y^{ql}_{S_1}]_{31} S_{1}^\dagger \bar{q}_{L, 3}^c i \sigma_2 l_{L,1} 
    + 
    [y^{eu}_{S_1}]_{13} S_{1}^\dagger \bar{e}_{R, 1}^c u_{R,3} 
    + \text{ h.c.} \,, \label{eq:LagrangianS1}
    \\
    -\mathcal{L}_{R_2} &= [y^{lu}_{R_2}]_{13} R_{2}^\dagger i \sigma_2 \bar{l}_{L, 1}^T u_{R,3} +
    [y^{eq}_{R_2}]_{13} R_{2}^\dagger  \bar{e}_{R, 1} q_{L,3} + \text{ h.c.} \,.
    \label{eq:LagrangianR2}
\end{align}
The tree-level matching conditions onto the five $eett$ coefficients entering our observables are then~\cite{deBlas:2017xtg}
\begin{align}
    [C_{lq}^{(1)}]_{1133} =& \frac{1}{4}\frac{[y^{ql}_{S_1}]_{31}^*[y^{ql}_{S_1}]_{31}}{M_{S_1}^2} \,, \nonumber \\
    [C_{lq}^{(3)}]_{1133} =& -\frac{1}{4}\frac{[y^{ql}_{S_1}]_{31}^*[y^{ql}_{S_1}]_{31}}{M_{S_1}^2} \,, \nonumber \\ \label{eq:eettTM}
    [C_{eu}]_{1133} =& \frac{1}{2}\frac{[y^{eu}_{S_1}]_{13}^*[y^{eu}_{S_1}]_{13}}{M_{S_1}^2} \,, \\
    [C_{lu}]_{1133} =& -\frac{1}{2}\frac{[y^{lu}_{R_2}]_{13}^*[y^{lu}_{R_2}]_{13}}{M_{R_2}^2} \,, \nonumber  \\
    [C_{qe}]_{3311} =& -\frac{1}{2}\frac{[y^{eq}_{R_2}]_{13}^*[y^{eq}_{R_2}]_{13}}{M_{R_2}^2}, \nonumber
\end{align}
i.e. we get a contribution in \textit{all} the $eett$ operators discussed in the previous sections.
This implies that, aside from the fixed relation between $\cC_{lq}^{(1)}$ and $\cC_{lq}^{(3)}$, these two states precisely mirror our EFT analysis, so that we can compare directly the EFT approach and an explicit model. 
In particular, the main difference will lie in the fact that we are able to include the finite one-loop matching terms at the high scale, and assess their impact compared to the RGE logarithms. 
These matching terms are qualitatively different than the finite EFT loop corrections computed e.g. in Refs.~\cite{Asteriadis:2024xts,Bellafronte:2025ubi} since they represent an additional contribution arising only within a specific UV model.
Notice also that on top of the operators listed above, both leptoquarks also have a non-zero matching condition onto the scalar and tensor semileptonic coefficients $\cC_{lequ}^{(1,3)}$. In our setup these are phenomenologically less relevant (cf. Section~\ref{sec:UVandRelatedStructures}) and we will ignore them in the following.

\begin{figure*}
    \centering
    \includegraphics[width=0.8\linewidth]{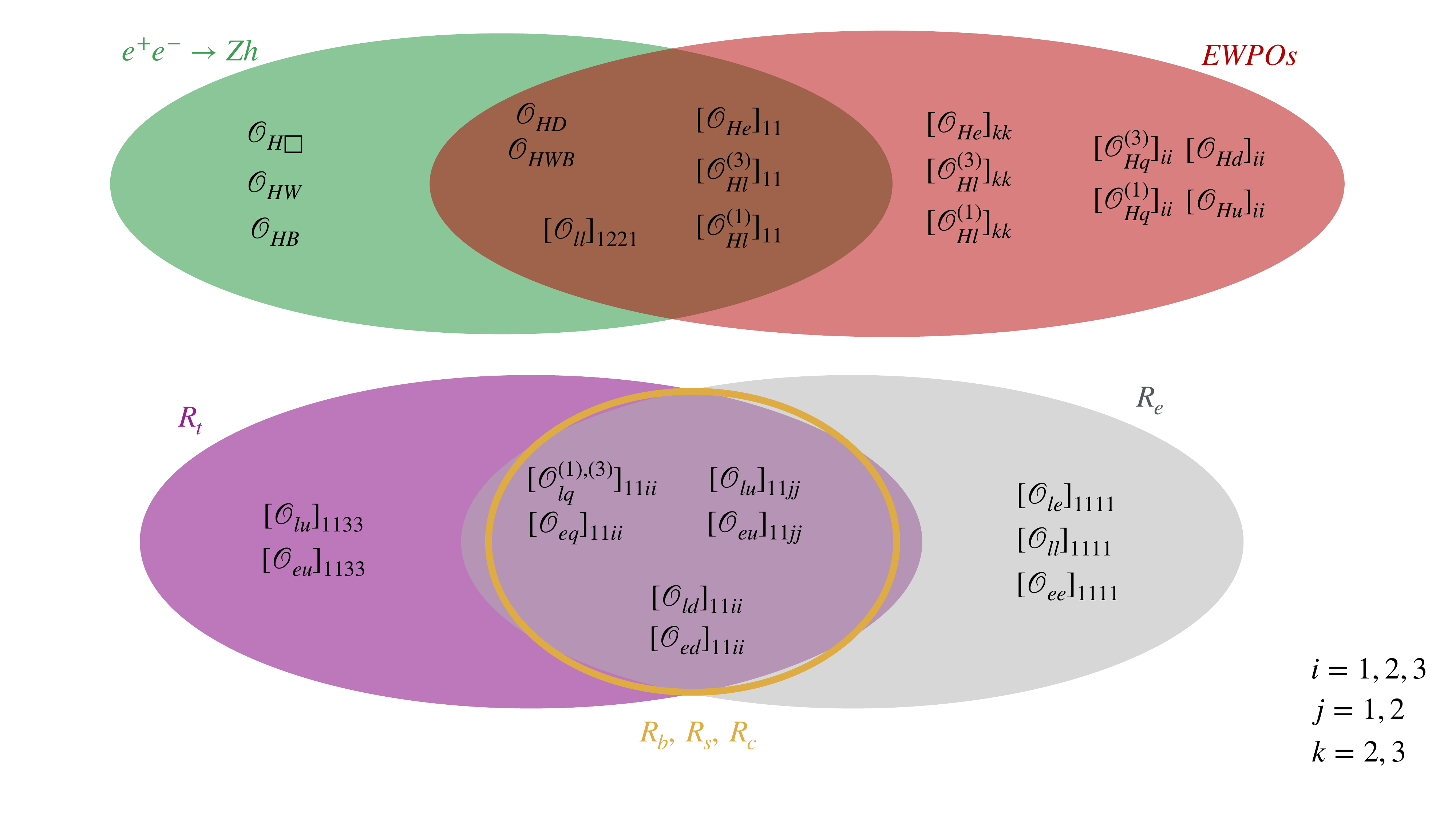}
    \caption{
    Operators included in our analysis when considering one-loop matching for the $S_1$ and $R_2$ leptoquarks, divided by observables in which they contribute. Only operators entering at tree-level in the respective observables are kept, as any other would correspond to a two-loop effect.}
    \label{fig:OperatorOneLoopMatching}
\end{figure*}

\begin{figure*}[t]
    \centering
   
    \begin{minipage}{0.47\textwidth}
        \centering
        \includegraphics[width=\linewidth]{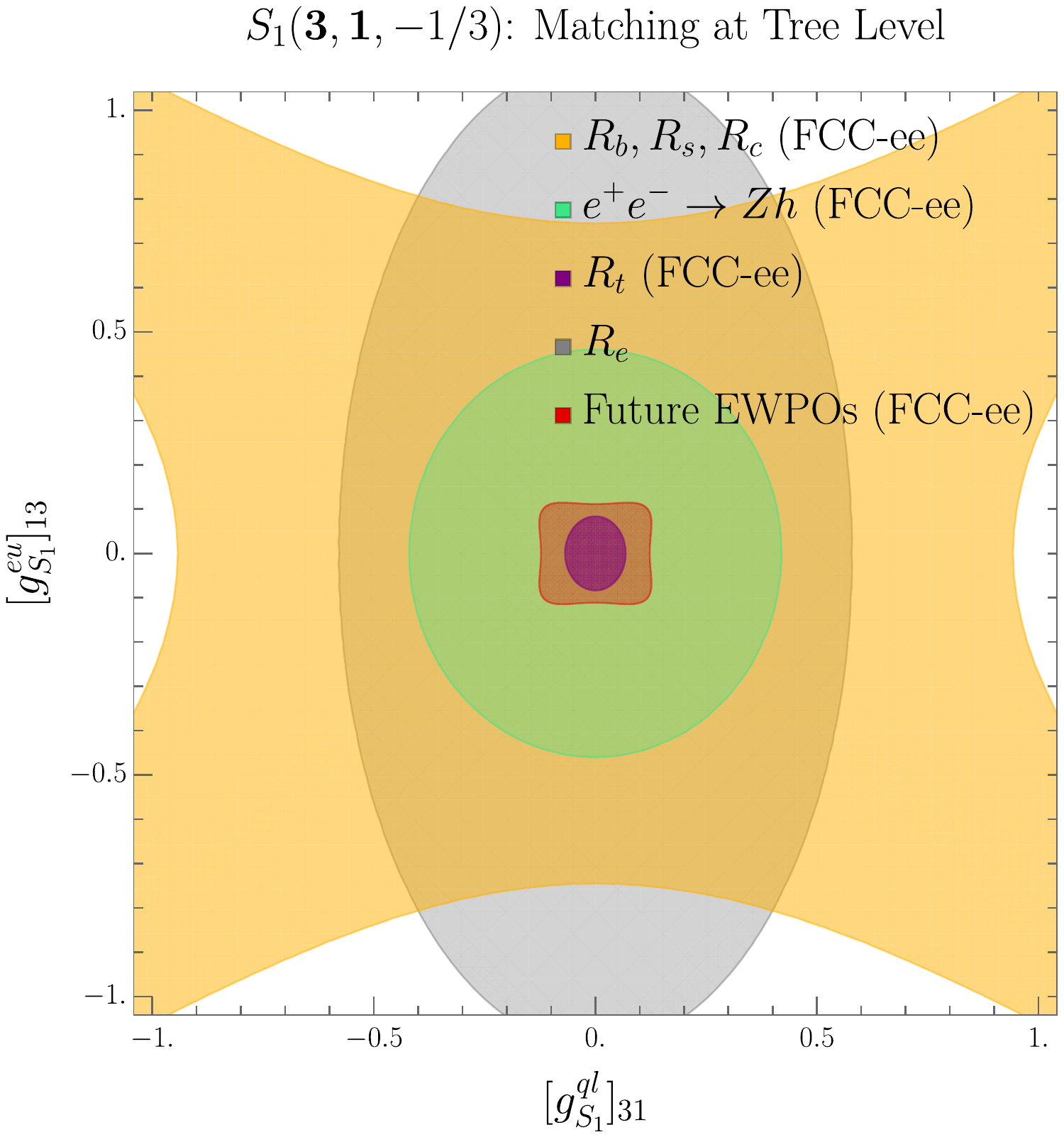}
    \end{minipage}
    \hfill \begin{minipage}{0.47\textwidth}
        \centering
        \includegraphics[width=\linewidth]{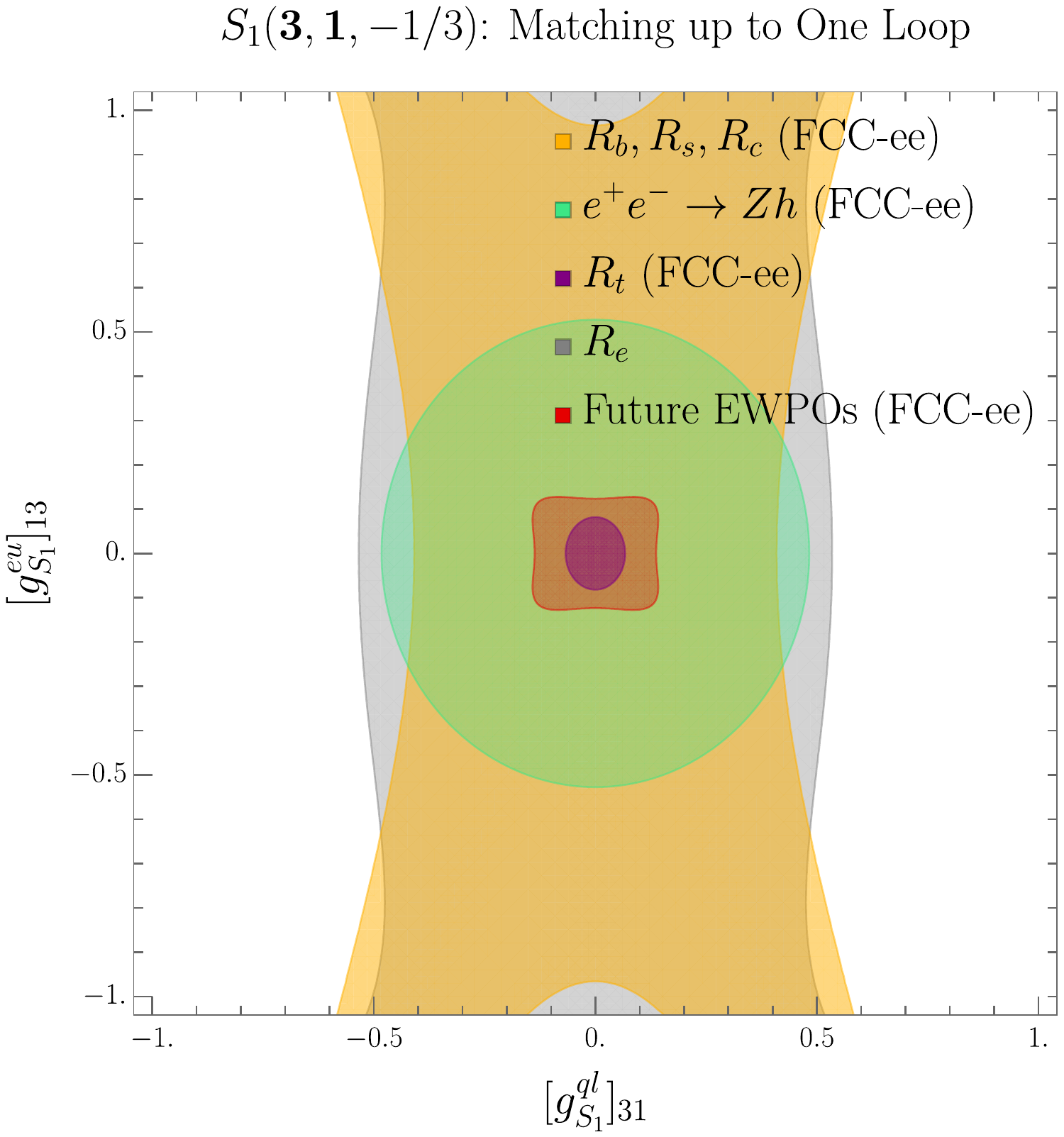}
    \end{minipage}

    \caption{2$\sigma$ C.L. allowed regions from different observables, for $S_1$ leptoquark only, with $M_{S_1}=1$\,TeV. 
    \textit{Left}: tree-level matching onto SMEFT at the high scale ($\Lambda=1$\,TeV). \textit{Right}: one-loop matching.}
    \label{fig:TMLMS1}
\end{figure*}

As a first example, we focus on the case of the $S_1$ alone, setting $[y^{lu}_{R_2}]_{13}=[y^{eq}_{R_2}]_{13}=0$.
Setting furthermore the mass of the leptoquark to $M_{S_1}=1$\,TeV \footnote{In the case of tree-level matching, this mass is irrelevant, since everything can be expressed as function of $g/M$, and the scale merely indicates the initial scale for the running. Going to one-loop matching, this relation is broken, and masses and couplings can in principle be disentangled. We still choose, for comparison with the EFT case, to keep $M=1$\,TeV. This could in principle be probed by direct searches at the LHC through leptoquark pair production. However we were not able to find any reference for a search with electrons and tops in the final state, as the literature typically focusses on fully first generation or purely third-generation couplings.},
we show in Fig.~\ref{fig:TMLMS1} the allowed regions for the two couplings.
The first feature that appears is that, combining all observables, both couplings are expected to be tightly constrained at FCC-ee.
In particular, EWPOs and $e^+e^-\to t\bar t$ play a crucial role, given the top-quark specific scenario considered.
Indeed, $e^+e^-\to b\bar b$ transitions appear to be less constraining in this scenario than in the EFT, which is due to the fact that the $\cC_{lq}^{(1)}-\cC_{lq}^{(3)}$ combination couples charged leptons to up-type quarks, and neutrinos to down-type quarks.
Nonetheless, a sensitivity to the ratios $R_q$ and $R_e$ is retained through loop effects.
This is due to the fact that the gauge coupling running breaks the tree-level relation.
The loop-matching conditions of two coefficents, where the first is modifying the observable $R_q$ and the second the $Zee$ coupling are~\cite{Gherardi:2020det}

\begin{align}
    \label{eq:Ceumatching}
    16 \pi^2 [\cC_{eu}]_{1122} &\approx 
    \frac{\, g^4}{135 \, M_{S_1}^2}+ + \frac{g^2 [y^{eu}_{S_1}]_{13}^*[y^{eu}_{S_1}]_{13}}{M_{S_1}^2} \left( \frac{5}{9} - \frac{9}{4} \log \left[ \frac{ M_{S_1}^2}{\mu^2}\right] \right) \\
   16 \pi^2 [\cC_{He}]_{11} &\approx 
   \  g^4 \frac{1}{180  M_{S_1}^2} 
    + \frac{g^2 [y^{eu}_{S_1}]_{13}^*[y^{eu}_{S_1}]_{13}}{M_{S_1}^2} \left( \frac{5}{12} - \frac{1}{3} \log \left[ \frac{ M_{S_1}^2}{\mu^2}\right] \right) \nonumber \\ \label{eq:CHematching}
    & - \frac{3 \ y_t^2 }{2}  \frac{[y^{eu}_{S_1}]_{13}^*[y^{eu}_{S_1}]_{13} }{M_{S_1}^2} \left( 1- \log \left[ \frac{ M_{S_1}^2}{\mu^2}\right]
   \right),
\end{align}
where we neglected all Yukawa couplings, except the top one~\footnote{In our analysis, we include only the finite terms of Eqs. \eqref{eq:Ceumatching} and \eqref{eq:CHematching}. The logarithms will be fully taken into account by the RGE. In other words, we choose the matching scale to be exactly the mass of the leptoquark, and evolve the couplings from that scale.}.
These matching conditions give an example of the different terms that might appear in such a computation.
In the first term in both equations we see a purely finite contribution depending on the gauge couplings only, which is coming from a correction to the SM gauge boson propagators.
The second terms contain both logs and finite terms obtained by closing the quark loop and coupling to the Higgs current through a $B$ field.
The logarithms are exactly the ones from the RGE, while the finite terms can only be obtained doing the full computation.
Similarly, the $y_t$-dependent terms in the equation of $[\cC_{He}]_{11}$ refer to diagrams as in Fig.~\ref{fig:FeynDiagramseett} (c), and the logarithms match exactly the one in Eq.~(\ref{eq:HemZ}), once the tree-level matching conditions are substituted (Eq.~\eqref{eq:eettTM}).
The impact of loop effects in these observables is further evident from the fact that the yellow and green regions change significantly once the finite one-loop matching terms are included (cf. Fig.~\ref{fig:TMLMS1} (right)), which we computed for all relevant coefficients with the help of {\tt SOLD}~\cite{Guedes:2023azv,Guedes:2024vuf} (see also Fig.~\ref{fig:OperatorOneLoopMatching} for the full collection of operators). 
Another sizable impact is seen in the case of the EWPO projections, for which the inclusion of the finite terms slightly relaxes the bound, in line with the findings in Ref.~\cite{Gargalionis:2024jaw}. All operators that enter thorugh loop-matching conditions in addition to the tree-level matching conditions of the $eett$ operators in the considered processes can be found in Fig.~\ref{fig:OperatorOneLoopMatching}.
The case of the $R_2$ leptoquark alone is qualitatively similar, and we refer to the appendix for explicit details.

In both cases, we find that, as expected, the extraction of the Higgs trilinear is only mildly affected by the new heavy states, yielding again  $\delta \kappa_3 = 0.169$, independent of the used order of matching. 
So even though we can see a change in the bounds in the couplings to the Leptoquarks, this is too small to affect the bounds on $\delta \kappa_3$. 

\begin{figure}[h]
    \centering
    \includegraphics[width=0.5\textwidth]{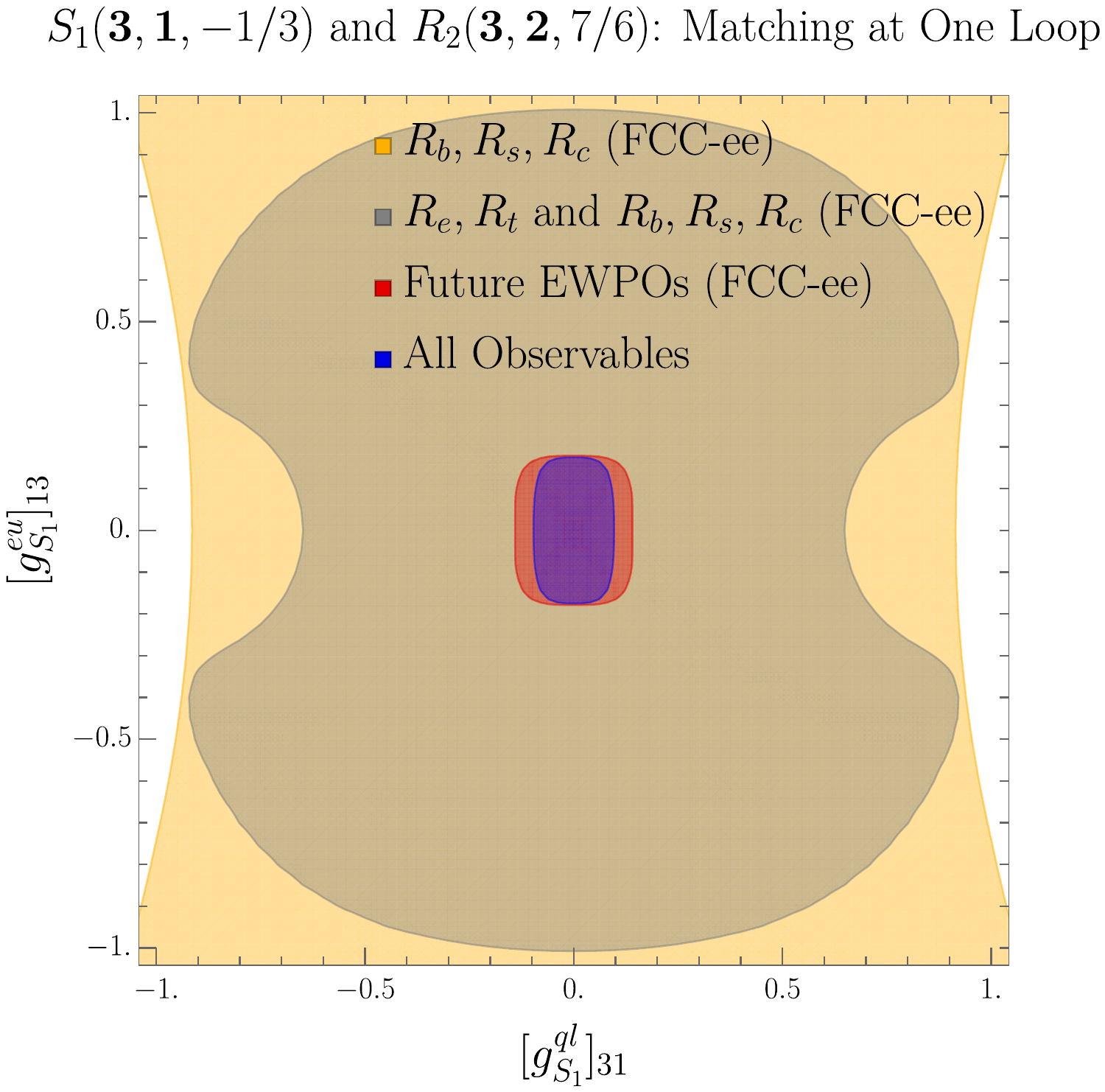}
    \caption{2$\sigma$ C.L. allowed regions from different observables for the $S_1$ couplings, marginalizing over the $R_2$ couplings. 
    The matching to SMEFT is done at  one-loop and the masses are set to $M_{S_1}= M_{R_2}=1$\,TeV.}
    \label{fig:MargS1}
\end{figure}

In Fig.~\ref{fig:MargS1} we show the allowed regions for the same $S_1$ couplings, but marginalising over the $R_2$ couplings.
This is to understand how switching on all of the $eett$ interactions affects a global fit in the explicit model case.
For brevity we only show the case of full one-loop matching.
The allowed regions, appear to be slightly larger, while $R_e$, $R_t$ and the Higgsstrahlung do not provide separate constraints anymore.
Nevertheless, the inclusion of all the observables allows for a good restriction of the parameter space.
Marginalising over all four couplings, the sensitivity to the Higgs trilinear is 
\begin{align}
    \delta \kappa_3 (S_1^{\rm tree} +R_2^{\rm tree}) &=0.172 \nonumber \\
    \delta \kappa_3 (S_1^{\rm loop} +R_2^{\rm loop}) &= 0.171 \,.
\end{align}
We can again observe nearly the same result for a one-loop and tree level matched model. As expected from the larger parameter space we get a slightly bigger value for $\delta \kappa_3$. Nevertheless, in both cases we do not see a flat direction and can restrict $\delta \kappa_3$ quite strongly. 

\section{Conclusion}

This letter addresses the question of the interplay of four-fermion contact interactions involving electrons and top-quarks, and the extraction of the Higgs self-coupling at a future $e^+e^-$ circular machine.
The interest arises from the fact that both, $eett$ interactions and the Higgs trilinear, enter the $e^+e^- \to Zh$ cross-section at one-loop~\cite{Asteriadis:2024xts,Bellafronte:2025ubi}.
Indeed, focussing only on $Zh$ production, the two separate effects are difficult to disentangle, and therefore can lead to a measurement which is subject to additional hypotheses about the UV, e.g. the one of no New Physics in $eett$ interactions.
Within SMEFT, however, the same interactions can be probed, both at tree-level and one-loop, with several other observables measured at current and future colliders, which therefore assist in constraining the parameter space.
Such interplays have been studied in recent works in the literature, focussing on global SMEFT fits featuring specific flavour symmetries~\cite{Maura:2025rpe}, or on specific models giving rise to a modification of the self-coupling~\cite{terHoeve:2025omu}.
Our work builds on and complements said results, by i) extending the phenomenological study to include prospected $ee\to ff$ measurements above the $Z$-pole, and ii) studying the impact on the Higgs trilinear not only within SMEFT, but also in explicit leptoquark models coupling to tops and electrons.
The first point leads to an improvent in the constraints on $eett$ operators, specifically removing a pseudo-flat direction in the fit presented in Ref.~\cite{Bellafronte:2025ubi}, and leading to bounds in the 10$\div$20\,TeV range even in the fully marginalised fit done with all $eett$ operators simultaneously.
This has a crucial impact on the extraction of the Higgs trilinear, since we showed that the four-fermion fit is largely independent on the Higgstrahlung cross-section.
As a consequence the measurement of $\kappa_3$ is largely unaffected by the presence of $eett$ interactions.

The two leptoquark models in the final section of the paper have been chosen as an example since they contribute to all $eett$ SMEFT operators. They can therefore replicate the EFT analysis, with the additional correlations which arise within a specific model, and the possibility of assessing the impact of the finite one-loop matching terms.
The results follow the ones of the EFT approach, with no major change in the precision of $\kappa_3$ once the one-loop terms at the matching scale are taken into account.

These findings once again confirm the incredible exploration power of a machine such as the FCC-ee.
Confronting it with different NP scenarios with couplings to electrons and top quarks, the extraction of $\kappa_3$ appears to be very robust, due to the rich and broad experimental programme, capable of providing many redundant and complementary constraints.

\section*{Ackowledgements}

We want to thank the authors of Ref.~\cite{Bellafronte:2025ubi} who, at the time of submission, shared with us a revised version of their study confirming the importance of FCC-ee high-energy runs to disentangle the eett interactions and the Higgs self-coupling.

The authors acknowledge funding from the Deutsche Forschungsgemeinschaft under Germany’s Excellence Strategy EXC 2121 “Quantum Universe” – 390833306, as well as from the grant 491245950.
LT’s research is funded by the Deutsche Forschungsgemeinschaft (DFG, German Research Foundation) - Projektnummer 417533893/GRK2575 “Rethinking Quantum Field Theory”

\appendix

\section{Constraints on $eett$ operators}

Here we report (Table~\ref{tab:Boundseett}) the numerical results for the bounds on electron-top coefficients appearing in the main text and in Fig.~\ref{fig:BoundsMargSep}, under inclusion of different sets of observables.
For details about the observables we refer back to the main text and to Table~\ref{tab:obs}.

{\renewcommand{\arraystretch}{1.4}
\begin{table*}
    \centering
    \resizebox{\textwidth}{!}{
    \begin{tabular}{c|cc|cc|cc|cc|cc}
             \multirow{2}{*}{[TeV]} & \multicolumn{2}{c|}{EWPO (current)}
            & \multicolumn{2}{c|}{+DY (HL-LHC)}
            & \multicolumn{2}{c|}{+EWPO (FCC-ee)}
            & \multicolumn{2}{c|}{+ $R_t$, $R_b$, $R_s$, $R_c$ }
            & \multicolumn{2}{c}{ + $e^+e^-\rightarrow Zh$ (all)} \\ 
           & sep & marg & sep & marg & sep & marg & sep & marg & sep & marg \\ \hline 
        $[C_{lq}^{(3)}]_{1133}$ & 1.9 & 1.6 & 2.1 & 1.6 & 17.6 & 14.4 & 40.3 & 18.0 & 40.3 & 18.0\\
        $[C_{lq}^{(1)}]_{1133}$ & 2.8 & - & 3.1 & 1.6 & 13.7 & 2.2 & 41.1 & 23.3 & 41.1 & 23.2\\
        $[C_{lu}]_{1133}$       & 2.7 & - & 2.7 & 1.4 & 13.3 & 2.2 & 19.3 & 9.2 & 19.3 & 9.4\\
        $[C_{qe}]_{3311}$       & 2.4 & - & 2.7 & 2.0 & 14.6 & 2.2  & 23.3 & 13.8 & 23.3 & 13.9\\
        $[C_{eu}]_{1133}$       & 2.4 & - & 2.4 & 1.5 & 14.2 & 2.1  & 20.3 & 9.4 & 20.3 & 9.6\\
    \end{tabular}}
    \caption{Bounds for the separate (setting all other SMEFT coefficients to zero) and marginalised over the other $eett$-WC for different observables (as in Fig. \ref{fig:BoundsMargSep}). All Bounds are given in TeV. See also Table~\ref{tab:obs} for the used observables.}
    \label{tab:Boundseett}
\end{table*}}

\section{Details about Drell--Yan}

\begin{figure}
    \centering
    \includegraphics[width=0.7\linewidth]{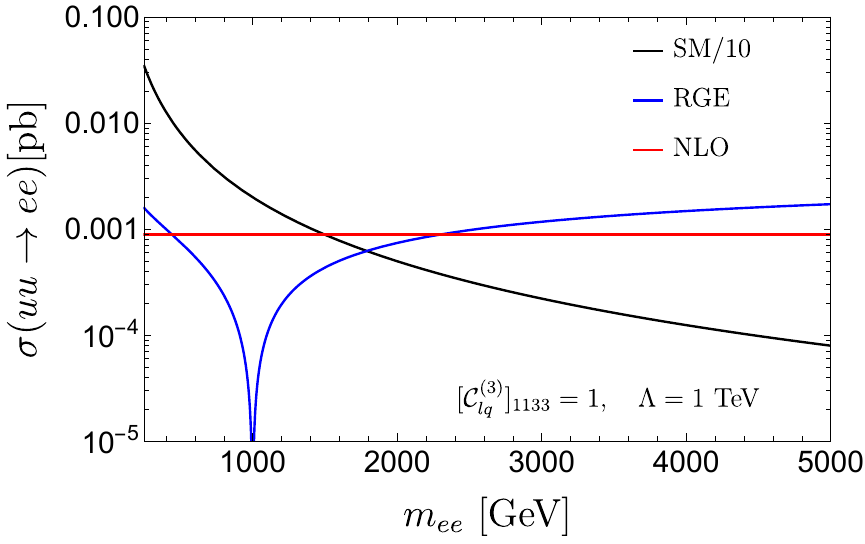}
    \caption{Comparison of different contributions to the partonic Drell--Yan cross-section. All coefficients except $[\cC_{lq}^{(3)}]_{1133}$ are set to zero at the scale $\Lambda$. The red (NLO) line is the contribution from the constant, finite terms presented in Ref.~\cite{Bellafronte:2025ubi} ($y_{ij}$ in our notation), while the blue indicates log-enhanced RG effects ($x_{ij}$). Similar plots can be obtained for other choices of Wilson coefficients.}
    \label{fig:DYxsec}
\end{figure}

The total partonic neutral Drell--Yan cross-section can be written schematically in SMEFT as
\begin{align}
    \label{eq:DYxsec}
    \hat\sigma_{qq\to \ell\ell}(\hat s) = \frac{a}{\hat s} \left(1 + \sum_i b_i\,  \cC_{H\psi\,,i}\right) + c_i\, \cC_{4\psi,i} + \mathcal{O}(\cC^2) \,,
\end{align}
where we have neglected the $Z$-boson mass (i.e. we are assuming $\hat s \gg m_Z^2$), and kept terms up to $1/\Lambda^2$ in the EFT expansion \footnote{Notice that for e.g. $bb\to \tau\tau$ the quadratic terms cannot be neglected, since they dominate the bounds, see Ref.~\cite{Allwicher:2022gkm} for a detailed discussion. In our analysis we always include the quadratic terms as well, we just do not include them here for brevity.}.
$a$ indicates the pure SM contribution, and can be computed to higher orders if needed.
However in Fig.~\ref{fig:DYxsec} we use the tree-level computation from {\tt HighPT}, giving $a = 5.08\times 10^{-5}$.
The SMEFT amplitude is also computed at tree-level.
The generic $\cC_{H\psi}$ indicates one or more operators of the class $\psi^2H^2 D$, which modify the $Z$-boson couplings, as described in the main text.
These contributions, given the $1/\hat s$ suppression, become less relevant in the tails, as they only represent a shift of the SM cross-section.
Moreover, constraints from EWPOs will always dominate and lead to a negligible contribution in Drell-Yan.
$\cC_{4\psi}$ indicates the tree-level contribution from semileptonic four-fermion operators.
For the process $u\bar u\to e^+e^-$, for example, the relevant operators are listed in the left column of Table~\ref{tab:DYnumbers}.
Their contribution is energy-enhanced with respect to the SM due to the point-like interaction, and leads to the well-known sensitivity in the tails.
Here we want to compare the contribution from RGE effects to the one from the finite terms first computed recently in Ref.~\cite{Bellafronte:2025ubi}, with focus on $eett$ operators.
Taking valence quarks in the initial state (and electrons in the final state), one can always write the relevant Wilson coefficient(s) at the scale $\hat s$ as
\begin{align}
    \label{eq:xy}
    [\cC_{4\psi,i}]_{1111} (\hat s) = \frac{1}{16\pi^2} \left(x_{ij} \log\frac{\sqrt{\hat s}}{\Lambda} + y_{ij}\right) \cC_{eett\,,j} (\Lambda) \,,
\end{align}
where $a_{ij}$ encodes the contribution from $\cC_{eett\,,j}$ (as defined in Eq.~(\ref{eq:eettcoeff})) in the anomalous dimension of a given four-fermion coefficient, and the $b_{ij}$ can be easily related to the $X_{0,i}$ computed in Ref.~\cite{Bellafronte:2025ubi}, by making use of the numerical coefficients in Table~\ref{tab:DYnumbers} and the relation
\begin{align}
    y_{ij} = \frac{20 G_F^2}{81\pi\Lambda^2} \frac{X_{0,j}}{c_i} \,.
\end{align}
We report the numerical values in Tables~\ref{tab:xij} and \ref{tab:yij}. In Fig.~\ref{fig:DYxsec} one can observe that, in most of the relevant range for the di-electron final state, the two contributions are comparable in size, with the dip at 1\,TeV invariant mass being due to the arbitrary choice of the initial scale in the logarithms.
This means that, for the particular choice of $\Lambda=1$\,TeV, RGE effects in bins around that invariant mass are negligible, as is to be expected. For  these reasons, we choose not to include the finite one-loop terms in our fits, as they would not drastically impact the numerical results. Indeed, even in the most extreme case of a coherent effect between finite and RGE effects, the overall scale of the bounds (from Drell-Yan only) is expected to change by an $\mathcal{O}(1)$ number. It should be further noted (see main text for details) that the overall impact of Drell-Yan in our global results is marginal.
\begin{table}[]
    \centering
    \renewcommand{\arraystretch}{1.3}
    \begin{tabular}{c|c}
        $\cC_{4\psi,i}$ & $c_i$ [$\times 10^{-10}$\,GeV$^{-2}$] \\ \hline
        $[\cC_{lq}^{(1)}]_{1111}$ & -5.11 \\
        $[\cC_{lq}^{(3)}]_{1111}$ & 5.11 \\
        $[\cC_{qe}]_{1111}$ & -0.94 \\
        $[\cC_{lu}]_{1111}$ & -1.89 \\
        $[\cC_{eu}]_{1111}$ & -3.78 \\
    \end{tabular}
    \caption{Numerical coefficients entering the cross-section for $u\bar u\to e^+e^-$ at tree-level and $\mathcal{O}(\Lambda^{-2})$, in the limit $\hat s \gg m_Z^2$ (cf. Eq.~(\ref{eq:DYxsec})).}
    \label{tab:DYnumbers}
\end{table}

\begin{table}[]
    \centering
    \renewcommand{\arraystretch}{1.3}
    \begin{tabular}{c|ccccc}
         & $[\cC_{lq}^{(1)}]_{1133}$ & $[\cC_{lq}^{(3)}]_{1133}$ & $[\cC_{eq}]_{3311}$ & $[\cC_{lu}]_{1133}$ & $[\cC_{eu}]_{1133}$ \\ \hline
        $[\cC_{lq}^{(1)}]_{1111}$ & 0.028 & 0. & 0. & 0.057 & 0. \\
        $[\cC_{lq}^{(3)}]_{1111}$ & 0. & 0.841 & 0. & 0. & 0. \\
        $[\cC_{eq}]_{1111}$ & 0. & 0. & 0.028 & 0. & 0.057 \\
        $[\cC_{lu}]_{1111}$ & 0.114 & 0. & 0. & 0.228 & 0. \\
        $[\cC_{eu}]_{1111}$ & 0. & 0. & 0.114 & 0. & 0.228
    \end{tabular}
    \caption{$x_{ij}$ coefficients entering the NLO Drell--Yan cross-section of $u\bar u \to e^+e^-$. See Eq.~(\ref{eq:xy}).}
    \label{tab:xij}
\end{table}

\begin{table}[]
    \centering
    \renewcommand{\arraystretch}{1.3}
    \begin{tabular}{c|ccccc}
         & $[\cC_{lq}^{(1)}]_{1133}$ & $[\cC_{lq}^{(3)}]_{1133}$ & $[\cC_{eq}]_{3311}$ & $[\cC_{lu}]_{1133}$ & $[\cC_{eu}]_{1133}$ \\ \hline
         $[\cC_{lq}^{(1)}]_{1111}$ & -0.701 & 0.059 & 0.074 & 0.118 & 0.149 \\
        $[\cC_{lq}^{(3)}]_{1111}$ & 0.701 & -0.059 & -0.074 & -0.118 & -0.149 \\
        $[\cC_{eq}]_{1111}$ & -3.798 & 0.319 & 0.404 & 0.637 & 0.807 \\
        $[\cC_{lu}]_{1111}$ & -1.899 & 0.159 & 0.202 & 0.319 & 0.404 \\
        $[\cC_{eu}]_{1111}$ & -0.95 & 0.08 & 0.101 & 0.159 & 0.202
    \end{tabular}
    \caption{$y_{ij}$ coefficients entering the NLO Drell--Yan cross-section of $u\bar u \to e^+e^-$. See Eq.~(\ref{eq:xy}).}
    \label{tab:yij}
\end{table}

\section{Constraints on the couplings of the $R_2$ Leptoquark}\label{app:LQ}

\begin{figure*}[t]
    \centering
   
    \begin{minipage}{0.48\textwidth}
        \centering
        \includegraphics[width=\linewidth]{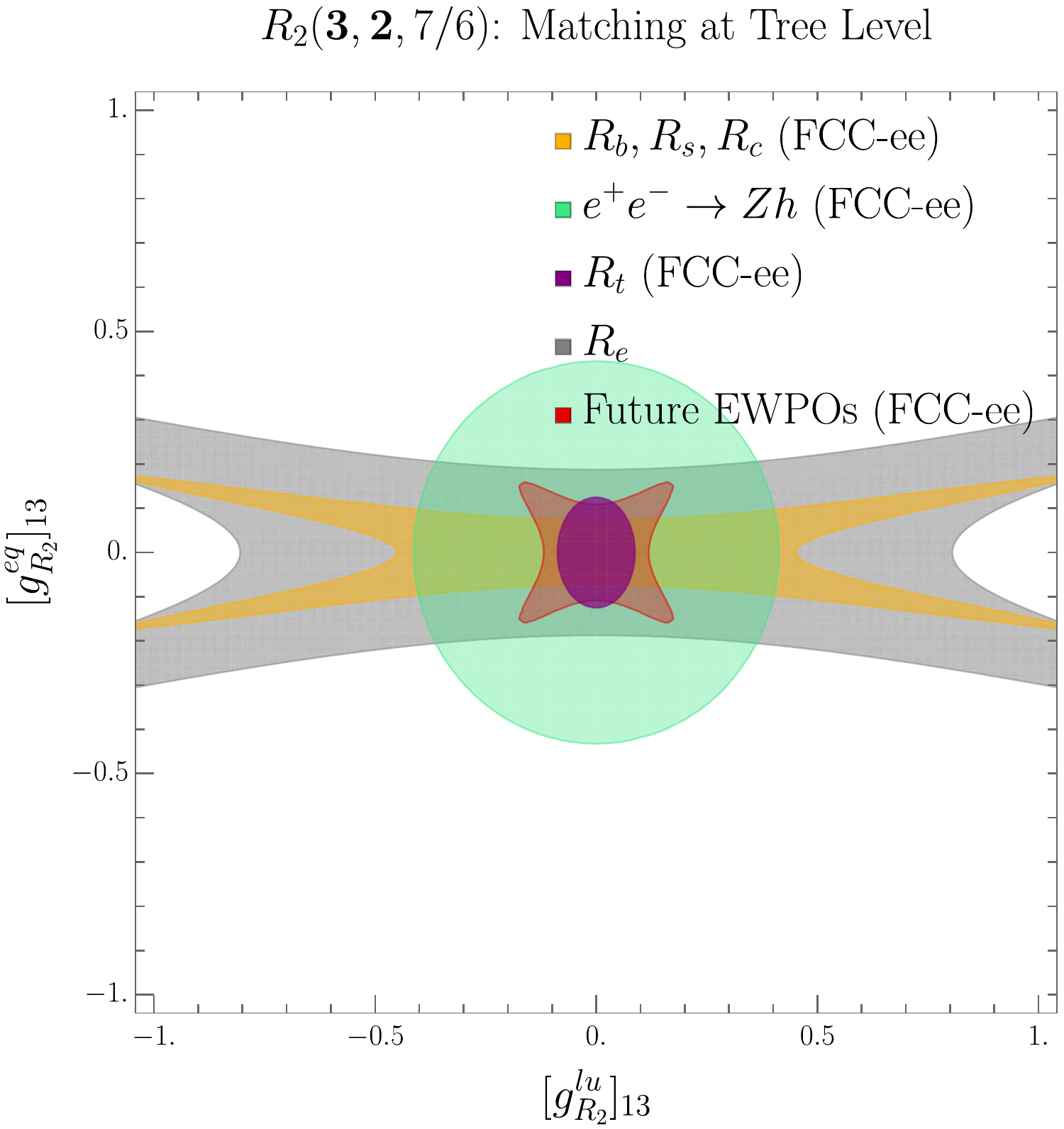}
    \end{minipage}
    \hfill \begin{minipage}{0.48\textwidth}
        \centering
        \includegraphics[width=\linewidth]{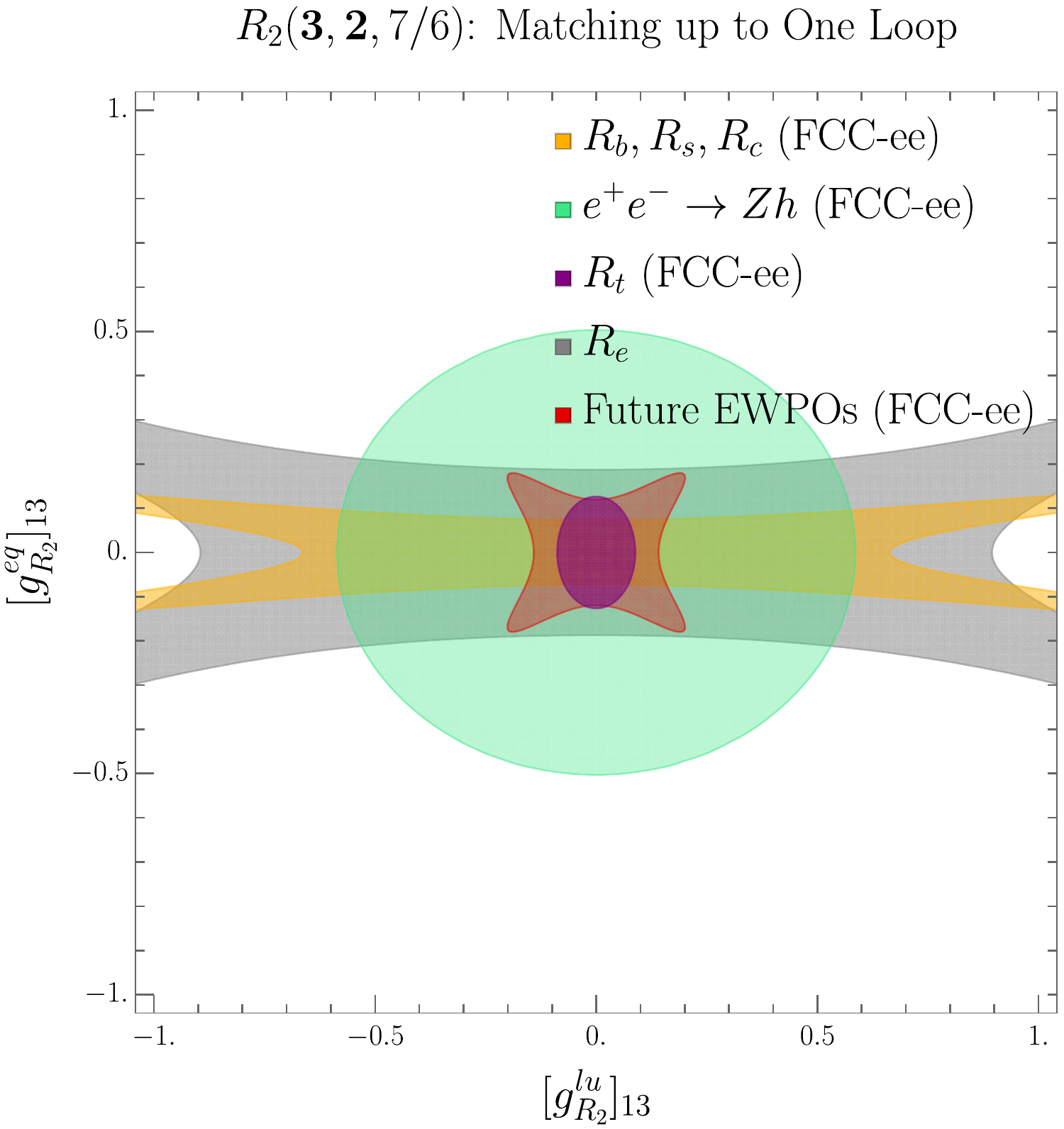}
    \end{minipage}

    \caption{The 2$\sigma$ C.L. allowed regions from different observables, for the one particle extension of the Leptoquark $R_2$ only, with $M_{R_2}=1$\,TeV. 
    The matching on the left is done at tree level and on the right at one-loop. 
    The couplings are defined in the Lagrangian of Eq.~(\ref{eq:LagrangianR2}).}
    \label{fig:TMLMR2}
\end{figure*}

\begin{figure}[h]
    \centering
    \includegraphics[width=0.5\linewidth]{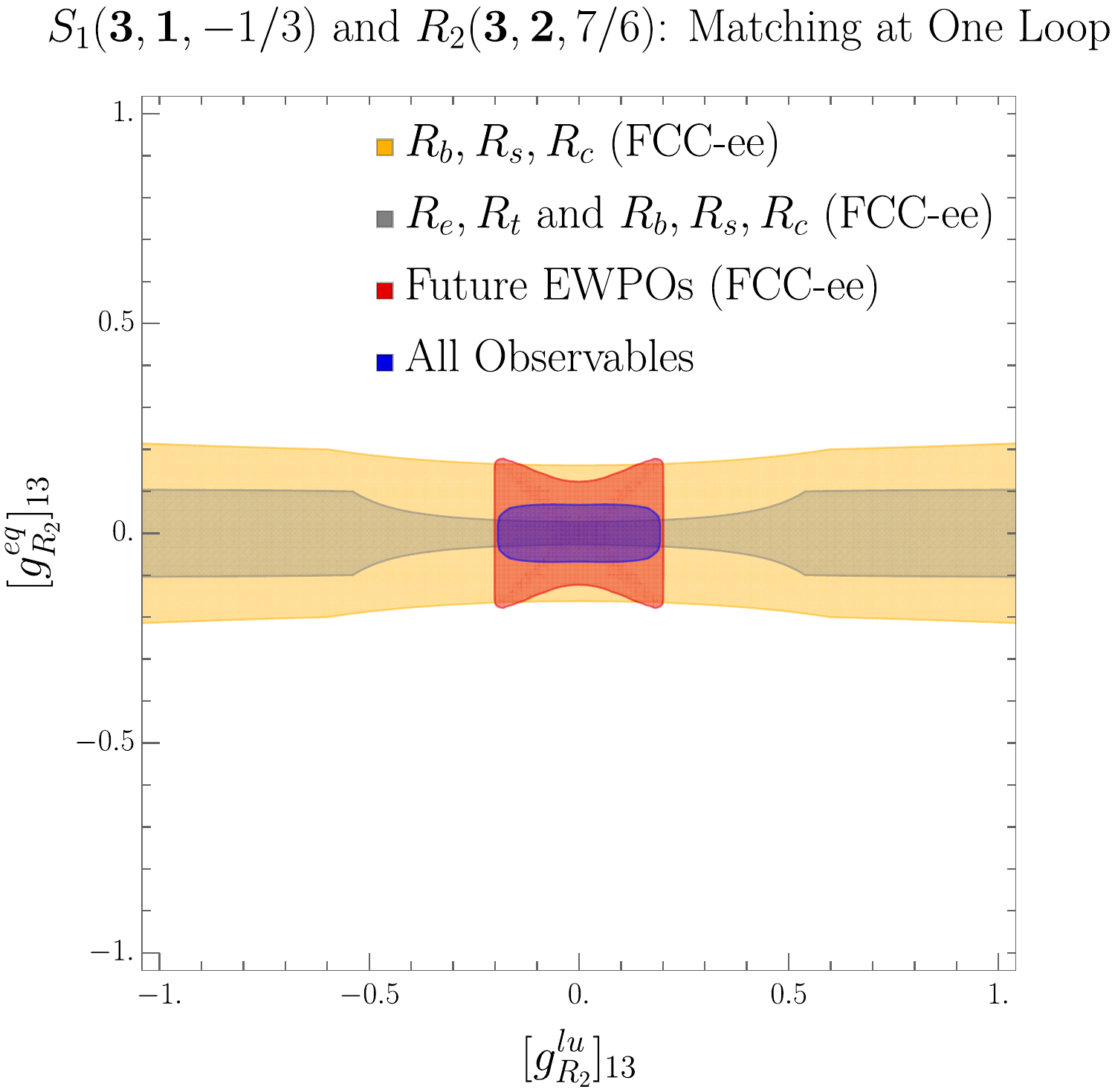}
    \caption{The 2$\sigma$ regions from different observables over the $R_2$ couplings. The regions are calculated by marginalizing over the $S_1$ couplings (see Eqs.~(\ref{eq:LagrangianR2}) and (\ref{eq:LagrangianS1})). The matching to SMEFT is done up to one-loop and the masses are set to $M_{R_2}=M_{S_1}=1$\,TeV.}
    \label{fig:MargR2}
\end{figure}

In addition to the figures in Section~\ref{sec:models} we show here the allowed regions for the $R_2$ leptoquark couplings. More specifically, Fig.~\ref{fig:TMLMR2} shows the bounds at tree-level and loop-level matching for the $R_2$ leptoquark alone, and Fig.~\ref{fig:MargR2} shows the bounds of the $R_2$ couplings, marginalized over the $S_1$ couplings.

\bibliographystyle{JHEP}
\bibliography{references}

\providecommand{\href}[2]{#2}\begingroup\raggedright\begin{thebibliography}{10}

\bibitem{CMS:2025ngq}
{\scshape CMS} collaboration, \emph{{Combination of searches for nonresonant
  Higgs boson pair production in proton-proton collisions at $\sqrt{s}$= 13
  TeV}},  \href{https://arxiv.org/abs/2510.07527}{{\ttfamily 2510.07527}}.

\bibitem{ATLAS:2024ish}
{\scshape ATLAS} collaboration, \emph{{Combination of Searches for Higgs Boson
  Pair Production in pp Collisions at s=13{\,}{\,}TeV with the ATLAS
  Detector}}, \href{https://doi.org/10.1103/PhysRevLett.133.101801}{\emph{Phys.
  Rev. Lett.} {\bfseries 133} (2024) 101801}
  [\href{https://arxiv.org/abs/2406.09971}{{\ttfamily 2406.09971}}].

\bibitem{Glover:1987nx}
E.W.N.~Glover and J.J.~van~der Bij, \emph{{HIGGS BOSON PAIR PRODUCTION VIA
  GLUON FUSION}},
  \href{https://doi.org/10.1016/0550-3213(88)90083-1}{\emph{Nucl. Phys. B}
  {\bfseries 309} (1988) 282}.

\bibitem{DiMicco:2019ngk}
J.~Alison et~al., \emph{{Higgs boson potential at colliders: Status and
  perspectives}}, \href{https://doi.org/10.1016/j.revip.2020.100045}{\emph{Rev.
  Phys.} {\bfseries 5} (2020) 100045}
  [\href{https://arxiv.org/abs/1910.00012}{{\ttfamily 1910.00012}}].

\bibitem{McCullough:2013rea}
M.~McCullough, \emph{{An Indirect Model-Dependent Probe of the Higgs
  Self-Coupling}},
  \href{https://doi.org/10.1103/PhysRevD.90.015001}{\emph{Phys. Rev. D}
  {\bfseries 90} (2014) 015001}
  [\href{https://arxiv.org/abs/1312.3322}{{\ttfamily 1312.3322}}].

\bibitem{DiVita:2017vrr}
S.~Di~Vita, G.~Durieux, C.~Grojean, J.~Gu, Z.~Liu, G.~Panico et~al., \emph{{A
  global view on the Higgs self-coupling at lepton colliders}},
  \href{https://doi.org/10.1007/JHEP02(2018)178}{\emph{JHEP} {\bfseries 02}
  (2018) 178} [\href{https://arxiv.org/abs/1711.03978}{{\ttfamily
  1711.03978}}].

\bibitem{Maltoni:2018ttu}
F.~Maltoni, D.~Pagani and X.~Zhao, \emph{{Constraining the Higgs self-couplings
  at e$^{+}$e$^{−}$ colliders}},
  \href{https://doi.org/10.1007/JHEP07(2018)087}{\emph{JHEP} {\bfseries 07}
  (2018) 087} [\href{https://arxiv.org/abs/1802.07616}{{\ttfamily
  1802.07616}}].

\bibitem{FCC:2025lpp}
{\scshape FCC} collaboration, \emph{{Future Circular Collider Feasibility Study
  Report: Volume 1, Physics, Experiments, Detectors}},
  \href{https://arxiv.org/abs/2505.00272}{{\ttfamily 2505.00272}}.

\bibitem{deBlas:2025gyz}
{\scshape Physics Preparatory Group} collaboration, \emph{{Physics Briefing
  Book: Input for the 2026 update of the European Strategy for Particle
  Physics}},  \href{https://arxiv.org/abs/2511.03883}{{\ttfamily 2511.03883}}.

\bibitem{Bellafronte:2025ubi}
L.~Bellafronte, S.~Dawson, P.P.~Giardino and H.~Liu, \emph{{Probing Top Quark -
  Electron Interactions at Future Colliders}},
  \href{https://arxiv.org/abs/2507.02039}{{\ttfamily 2507.02039}}.

\bibitem{Barbieri:2007bh}
R.~Barbieri, B.~Bellazzini, V.S.~Rychkov and A.~Varagnolo, \emph{{The Higgs
  boson from an extended symmetry}},
  \href{https://doi.org/10.1103/PhysRevD.76.115008}{\emph{Phys. Rev. D}
  {\bfseries 76} (2007) 115008}
  [\href{https://arxiv.org/abs/0706.0432}{{\ttfamily 0706.0432}}].

\bibitem{Elias-Miro:2013eta}
J.~Elias-Mir{\'o}, C.~Grojean, R.S.~Gupta and D.~Marzocca, \emph{{Scaling and
  tuning of EW and Higgs observables}},
  \href{https://doi.org/10.1007/JHEP05(2014)019}{\emph{JHEP} {\bfseries 05}
  (2014) 019} [\href{https://arxiv.org/abs/1312.2928}{{\ttfamily 1312.2928}}].

\bibitem{Contino:2013kra}
R.~Contino, M.~Ghezzi, C.~Grojean, M.~Muhlleitner and M.~Spira,
  \emph{{Effective Lagrangian for a light Higgs-like scalar}},
  \href{https://doi.org/10.1007/JHEP07(2013)035}{\emph{JHEP} {\bfseries 07}
  (2013) 035} [\href{https://arxiv.org/abs/1303.3876}{{\ttfamily 1303.3876}}].

\bibitem{Allwicher:2023shc}
L.~Allwicher, C.~Cornella, G.~Isidori and B.A.~Stefanek, \emph{{New physics in
  the third generation. A comprehensive SMEFT analysis and future prospects}},
  \href{https://doi.org/10.1007/JHEP03(2024)049}{\emph{JHEP} {\bfseries 03}
  (2024) 049} [\href{https://arxiv.org/abs/2311.00020}{{\ttfamily
  2311.00020}}].

\bibitem{Maura:2024zxz}
V.~Maura, B.A.~Stefanek and T.~You, \emph{{Accuracy complements energy:
  electroweak precision tests at Tera-Z}},
  \href{https://doi.org/10.1007/JHEP10(2025)022}{\emph{JHEP} {\bfseries 10}
  (2025) 022} [\href{https://arxiv.org/abs/2412.14241}{{\ttfamily
  2412.14241}}].

\bibitem{terHoeve:2025omu}
J.~ter Hoeve, L.~Mantani, J.~Rojo, A.N.~Rossia and E.~Vryonidou, \emph{{Higgs
  trilinear coupling in the standard model effective field theory at the high
  luminosity LHC and the FCC-ee}},
  \href{https://doi.org/10.1103/qtz8-bkd4}{\emph{Phys. Rev. D} {\bfseries 112}
  (2025) 013008} [\href{https://arxiv.org/abs/2504.05974}{{\ttfamily
  2504.05974}}].

\bibitem{Maura:2025rpe}
V.~Maura, B.A.~Stefanek and T.~You, \emph{{Higgs Self-Coupling at the Future
  Circular e+e- Collider}},
  \href{https://doi.org/10.1103/wjcy-1qk6}{\emph{Phys. Rev. Lett.} {\bfseries
  135} (2025) 141802} [\href{https://arxiv.org/abs/2503.13719}{{\ttfamily
  2503.13719}}].

\bibitem{Asteriadis:2024xts}
K.~Asteriadis, S.~Dawson, P.P.~Giardino and R.~Szafron, \emph{{e$^{+}$e$^{−}$
  {\textrightarrow} ZH process in the SMEFT beyond leading order}},
  \href{https://doi.org/10.1007/JHEP02(2025)162}{\emph{JHEP} {\bfseries 02}
  (2025) 162} [\href{https://arxiv.org/abs/2409.11466}{{\ttfamily
  2409.11466}}].

\bibitem{Grzadkowski:2010es}
B.~Grzadkowski, M.~Iskrzynski, M.~Misiak and J.~Rosiek, \emph{{Dimension-Six
  Terms in the Standard Model Lagrangian}},
  \href{https://doi.org/10.1007/JHEP10(2010)085}{\emph{JHEP} {\bfseries 10}
  (2010) 085} [\href{https://arxiv.org/abs/1008.4884}{{\ttfamily 1008.4884}}].

\bibitem{Allwicher:2022mcg}
L.~Allwicher, D.A.~Faroughy, F.~Jaffredo, O.~Sumensari and F.~Wilsch,
  \emph{{HighPT: A tool for~ high-$p_T$ Drell-Yan tails beyond the standard
  model}}, \href{https://doi.org/10.1016/j.cpc.2023.108749}{\emph{Comput. Phys.
  Commun.} {\bfseries 289} (2023) 108749}
  [\href{https://arxiv.org/abs/2207.10756}{{\ttfamily 2207.10756}}].

\bibitem{Breso-Pla:2021qoe}
V.~Bres{\'o}-Pla, A.~Falkowski and M.~Gonz{\'a}lez-Alonso, \emph{{A$_{FB}$ in
  the SMEFT: precision Z physics at the LHC}},
  \href{https://doi.org/10.1007/JHEP08(2021)021}{\emph{JHEP} {\bfseries 08}
  (2021) 021} [\href{https://arxiv.org/abs/2103.12074}{{\ttfamily
  2103.12074}}].

\bibitem{Greljo:2024ytg}
A.~Greljo, H.~Tiblom and A.~Valenti, \emph{{New physics through flavor tagging
  at FCC-ee}},
  \href{https://doi.org/10.21468/SciPostPhys.18.5.152}{\emph{SciPost Phys.}
  {\bfseries 18} (2025) 152}
  [\href{https://arxiv.org/abs/2411.02485}{{\ttfamily 2411.02485}}].

\bibitem{Allwicher:2022gkm}
L.~Allwicher, D.A.~Faroughy, F.~Jaffredo, O.~Sumensari and F.~Wilsch,
  \emph{{Drell-Yan tails beyond the Standard Model}},
  \href{https://doi.org/10.1007/JHEP03(2023)064}{\emph{JHEP} {\bfseries 03}
  (2023) 064} [\href{https://arxiv.org/abs/2207.10714}{{\ttfamily
  2207.10714}}].

\bibitem{Grunwald:2025kot}
C.~Grunwald, G.~Hiller, K.~Kr{\"o}ninger and L.~Nollen, \emph{{Beyond
  Universality: Probing Lepton Flavor in the SMEFT}},
  \href{https://arxiv.org/abs/2511.07089}{{\ttfamily 2511.07089}}.

\bibitem{Jenkins:2013wua}
E.E.~Jenkins, A.V.~Manohar and M.~Trott, \emph{{Renormalization Group Evolution
  of the Standard Model Dimension Six Operators II: Yukawa Dependence}},
  \href{https://doi.org/10.1007/JHEP01(2014)035}{\emph{JHEP} {\bfseries 01}
  (2014) 035} [\href{https://arxiv.org/abs/1310.4838}{{\ttfamily 1310.4838}}].

\bibitem{Crivellin:2020mjs}
A.~Crivellin, C.~Greub, D.~M{\"u}ller and F.~Saturnino, \emph{{Scalar
  Leptoquarks in Leptonic Processes}},
  \href{https://doi.org/10.1007/JHEP02(2021)182}{\emph{JHEP} {\bfseries 02}
  (2021) 182} [\href{https://arxiv.org/abs/2010.06593}{{\ttfamily
  2010.06593}}].

\bibitem{Davoudiasl:2023huk}
H.~Davoudiasl and P.P.~Giardino, \emph{{Electron g-2 foreshadowing discoveries
  at FCC-ee}}, \href{https://doi.org/10.1103/PhysRevD.109.075037}{\emph{Phys.
  Rev. D} {\bfseries 109} (2024) 075037}
  [\href{https://arxiv.org/abs/2311.12112}{{\ttfamily 2311.12112}}].

\bibitem{Allwicher:2025mmc}
L.~Allwicher, M.~McCullough, S.~Renner, D.~Rocha and B.~Smith, \emph{{The Price
  of a Large Electron Yukawa Modification}},
  \href{https://arxiv.org/abs/2511.02642}{{\ttfamily 2511.02642}}.

\bibitem{Ethier:2021bye}
{\scshape SMEFiT} collaboration, \emph{{Combined SMEFT interpretation of Higgs,
  diboson, and top quark data from the LHC}},
  \href{https://doi.org/10.1007/JHEP11(2021)089}{\emph{JHEP} {\bfseries 11}
  (2021) 089} [\href{https://arxiv.org/abs/2105.00006}{{\ttfamily
  2105.00006}}].

\bibitem{Celada:2024mcf}
E.~Celada, T.~Giani, J.~ter Hoeve, L.~Mantani, J.~Rojo, A.N.~Rossia et~al.,
  \emph{{Mapping the SMEFT at high-energy colliders: from LEP and the (HL-)LHC
  to the FCC-ee}}, \href{https://doi.org/10.1007/JHEP09(2024)091}{\emph{JHEP}
  {\bfseries 09} (2024) 091}
  [\href{https://arxiv.org/abs/2404.12809}{{\ttfamily 2404.12809}}].

\bibitem{Allwicher:2025bub}
L.~Allwicher, G.~Isidori and M.~Pesut, \emph{{Flavored circular collider:
  cornering New Physics at FCC-ee via flavor-changing processes}},
  \href{https://doi.org/10.1140/epjc/s10052-025-14359-8}{\emph{Eur. Phys. J. C}
  {\bfseries 85} (2025) 631}
  [\href{https://arxiv.org/abs/2503.17019}{{\ttfamily 2503.17019}}].

\bibitem{Bordone:2025cde}
M.~Bordone, C.~Cornella and J.~Davighi, \emph{{Precision tests in $b\rightarrow
  s\ell ^+\ell ^-$ ($\ell =e,\mu $) at FCC-ee}},
  \href{https://doi.org/10.1140/epjc/s10052-025-14696-8}{\emph{Eur. Phys. J. C}
  {\bfseries 85} (2025) 995}
  [\href{https://arxiv.org/abs/2503.22635}{{\ttfamily 2503.22635}}].

\bibitem{Fuentes-Martin:2020zaz}
J.~Fuentes-Martin, P.~Ruiz-Femenia, A.~Vicente and J.~Virto, \emph{{DsixTools
  2.0: The Effective Field Theory Toolkit}},
  \href{https://doi.org/10.1140/epjc/s10052-020-08778-y}{\emph{Eur. Phys. J. C}
  {\bfseries 81} (2021) 167}
  [\href{https://arxiv.org/abs/2010.16341}{{\ttfamily 2010.16341}}].

\bibitem{cornetgomez2025futurecolliderconstraintstopquark}
F.~Cornet-Gomez, V.~Miralles, M.~Miralles~L{\'o}pez, M.~Moreno~Ll{\'a}cer and
  M.~Vos, \emph{Future collider constraints on top-quark operators},  2025.

\bibitem{Dorsner:2016wpm}
I.~Dor{\v{s}}ner, S.~Fajfer, A.~Greljo, J.F.~Kamenik and N.~Ko{\v{s}}nik,
  \emph{{Physics of leptoquarks in precision experiments and at particle
  colliders}}, \href{https://doi.org/10.1016/j.physrep.2016.06.001}{\emph{Phys.
  Rept.} {\bfseries 641} (2016) 1}
  [\href{https://arxiv.org/abs/1603.04993}{{\ttfamily 1603.04993}}].

\bibitem{deBlas:2017xtg}
J.~de~Blas, J.C.~Criado, M.~Perez-Victoria and J.~Santiago, \emph{{Effective
  description of general extensions of the Standard Model: the complete
  tree-level dictionary}},
  \href{https://doi.org/10.1007/JHEP03(2018)109}{\emph{JHEP} {\bfseries 03}
  (2018) 109} [\href{https://arxiv.org/abs/1711.10391}{{\ttfamily
  1711.10391}}].

\bibitem{Gherardi:2020det}
V.~Gherardi, D.~Marzocca and E.~Venturini, \emph{{Matching scalar leptoquarks
  to the SMEFT at one loop}},
  \href{https://doi.org/10.1007/JHEP07(2020)225}{\emph{JHEP} {\bfseries 07}
  (2020) 225} [\href{https://arxiv.org/abs/2003.12525}{{\ttfamily
  2003.12525}}].

\bibitem{Guedes:2023azv}
G.~Guedes, P.~Olgoso and J.~Santiago, \emph{{Towards the one loop IR/UV
  dictionary in the SMEFT: One loop generated operators from new scalars and
  fermions}},
  \href{https://doi.org/10.21468/SciPostPhys.15.4.143}{\emph{SciPost Phys.}
  {\bfseries 15} (2023) 143}
  [\href{https://arxiv.org/abs/2303.16965}{{\ttfamily 2303.16965}}].

\bibitem{Guedes:2024vuf}
G.~Guedes and P.~Olgoso, \emph{{From the EFT to the UV: the complete SMEFT
  one-loop dictionary}},  \href{https://arxiv.org/abs/2412.14253}{{\ttfamily
  2412.14253}}.

\bibitem{Gargalionis:2024jaw}
J.~Gargalionis, J.~Quevillon, P.N.H.~Vuong and T.~You, \emph{{Linear Standard
  Model extensions in the SMEFT at one loop and Tera-Z}},
  \href{https://doi.org/10.1007/JHEP07(2025)136}{\emph{JHEP} {\bfseries 07}
  (2025) 136} [\href{https://arxiv.org/abs/2412.01759}{{\ttfamily
  2412.01759}}].

\end{thebibliography}\endgroup

\end{document}